\documentclass[a4paper,fleqn]{cas-dc}

\usepackage[utf8]{inputenc}

\usepackage[sorting=none,maxnames=2,minnames=1,maxbibnames=99,style=numeric-comp,giveninits=true]{biblatex} 
\makeatletter                                                                   
\newlength{\bibsep}{\@listi \global\bibsep\itemsep \global\advance\bibsep by\parsep}
\makeatother
\bibliography{cas-refs.bib}
\usepackage{siunitx}
\usepackage[version=4]{mhchem} 

\usepackage{hyperref}
\hypersetup{
	colorlinks=true,
	linkcolor=black,
	filecolor=black,      
	urlcolor=blue,
	citecolor=black,
	}
\usepackage{url}
\usepackage{doi}

\usepackage[none]{hyphenat}
\usepackage{microtype}
\usepackage[nameinlink]{cleveref}
\usepackage{cuted}

\DeclareUnicodeCharacter{2212}{\textendash}

\def\tsc#1{\csdef{#1}{\textsc{\lowercase{#1}}\xspace}}
\tsc{WGM}
\tsc{QE}
\tsc{EP}
\tsc{PMS}
\tsc{BEC}
\tsc{DE}

\begin{document}

\let\WriteBookmarks\relax
\def\floatpagepagefraction{1}
\def\textpagefraction{.001}

\shorttitle{A matter of performance \& criticality}

\shortauthors{W. Liu \textit{et al.}}

\title [mode = title]{A matter of performance \& criticality: a review of rare-earth-based magnetocaloric intermetallic compounds for hydrogen liquefaction}                 

\author[1]{Wei Liu}[ orcid=0000-0002-6094-5036]
\cormark[1]
\ead{wei.liu@tu-darmstadt.de}
\author[2]{Tino Gottschall}
\author[1]{Franziska Scheibel}
\author[2]{Eduard Bykov}
\author[1]{Alex Aubert}
\author[1]{Nuno Fortunato}
\author[1]{Benedikt Beckmann}
\author[1]{Allan M. Döring}
\author[1]{Hongbin Zhang}
\author[1]{Konstantin Skokov}
\author[1]{Oliver Gutfleisch}

\affiliation[1]{organization={TU Darmstadt},
    addressline={Peter-Grünberg-Str. 16}, 
    city={Darmstadt},
    citysep={}, 
    postcode={64287}, 
    country={Germany}
    }

\affiliation[2]{organization={Dresden High Magnetic Field Laboratory (HLD-EMFL), Helmholtz-Zentrum Dresden-Rossendorf},
    addressline={Bautzner Landstraße 400}, 
    city={Dresden},
    citysep={}, 
    postcode={01328}, 
    country={Germany}
    }

\cortext[cor1]{Corresponding Author}

\begin{abstract}
The low efficiency of conventional liquefaction technologies based on the Joule-Thomson expansion makes liquid hydrogen currently not attractive enough for large-scale energy-related technologies that are important for the transition to a carbon-neutral society. Magnetocaloric hydrogen liquefaction has great potential to achieve higher efficiency and is therefore a crucial enabler for affordable liquid hydrogen. Cost-effective magnetocaloric materials with large magnetic entropy and adiabatic temperature changes in the temperature range of 77 $\sim$ 20 K under commercially practicable magnetic fields are the foundation for the success of magnetocaloric hydrogen liquefaction. Heavy rare-earth-based magnetocaloric intermetallic compounds generally show excellent magnetocaloric performances, but the heavy rare-earth elements (Gd, Tb, Dy, Ho, Er, and Tm) are highly critical in resources. Yttrium and light rare-earth elements (La, Ce, Pr, and Nd) are relatively abundant, but their alloys generally show less excellent magnetocaloric properties. A dilemma appears: higher performance or lower criticality? In this review, we study how cryogenic temperature influences magnetocaloric performance by first reviewing heavy rare-earth-based intermetallic compounds. Next, we look at light rare-earth-based, "mixed" rare-earth-based, and Gd-based intermetallic compounds with the nature of the phase transition order taken into consideration, and summarize ways to resolve the dilemma.
\end{abstract}


\maketitle

{\small{\tableofcontents}}


\section{Introduction}\label{introduction}

Hydrogen energy will play an essential role in the transition to a carbon-neutral society \cite{Liu.2022,Glenk.2019,Zuttel.2010,Pagliaro.2010}. Liquid hydrogen is of great importance for efficient storage and transportation of hydrogen \cite{Zuttel.2010,Aasadnia.2018,Durbin.2013,Wijayanta.2019,Aziz.2021}. Conventional hydrogen liquefaction technologies are based on the Joule-Thomson expansion, a thermodynamic process in which a gas or liquid undergoes changes in temperature and pressure during adiabatic expansion. Currently, the best conventional industrial hydrogen liquefier can only achieve an efficiency of 23\% \cite{Aasadnia.2018}. The low efficiency of the conventional technologies makes liquid hydrogen uneconomical for large-scale energy-related applications such as hydrogen-powered vehicles and hydrogen energy storage \cite{Durbin.2013}.\par

\begin{figure}[htbp]
    \centering
    \includegraphics[width=\linewidth]{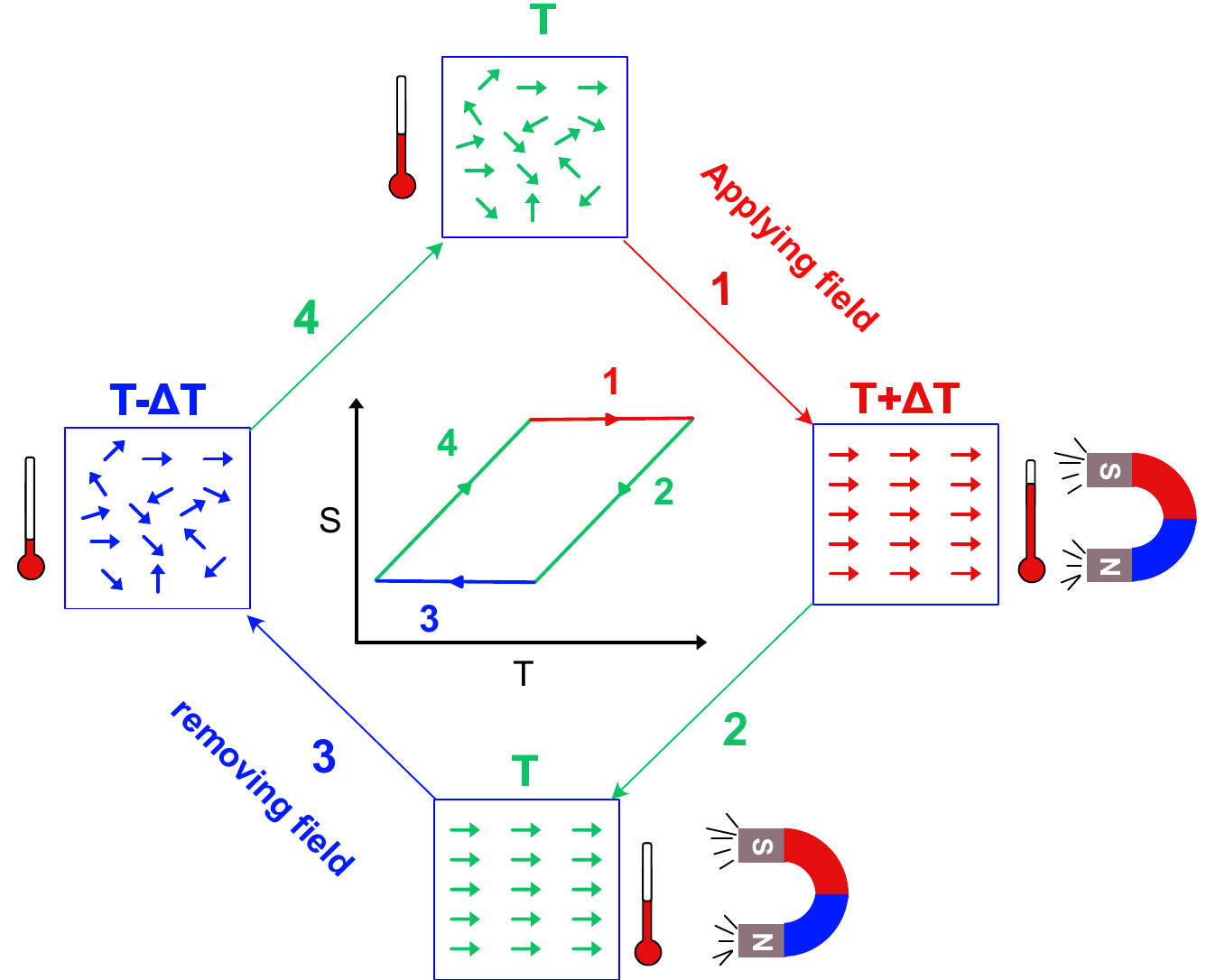}
    \caption{Illustration of a magnetic cooling cycle. $T$ is the temperature, $\Delta T$ is the temperature change, and $S$ is the total entropy.}
    \label{fig1}
\end{figure}

Magnetocaloric (MC) gas liquefaction is an emerging technology based on the magnetocaloric effect (MCE) \cite{Liu.2022,beckmann_dissipation_2023,bykov_magnetocaloric_2021,Terada.2021,Yang.2023,barclay_propane_2019,archipley_methane_2022}. The working principle of a magnetic cooling cycle is illustrated in \textbf{\Cref{fig1}}. First, the MC material is magnetized by applying a magnetic field under adiabatic conditions. The total entropy $S$ remains constant, while the magnetic entropy $S_m$ decreases. The lattice entropy $S_l$ increases to compensate for the decrease in $S_m$, resulting in an increase in the temperature of the MC material. The heat generated during the adiabatic magnetization process is then expelled by the exchange fluid. Then, the MC material is demagnetized by removing the field under adiabatic conditions, leading to a decrease in the temperature of the MC material. Finally, the heat from the environment is absorbed by the MC materials until the system returns to its initial state.

As an alternative to conventional hydrogen liquefaction technologies, MC hydrogen liquefaction demonstrates great potential to achieve higher efficiency \cite{Matsumoto.2009,kamiya_active_2022,Kitanovski.2020,moya2014caloric}. Feng \textit{et al.} showed an efficiency of more than 60\% in their model \cite{Feng.2020}. It should be emphasized that no evaluation of the efficiency of a real MC hydrogen liquefaction device has been conducted yet. Nevertheless, this theoretical efficiency is still encouraging considering that it is much higher than that of the best conventional industrial hydrogen liquefier mentioned above. The potential high efficiency of MC hydrogen liquefaction has attracted worldwide attention, leading to a rapidly growing interest in this emerging technology for its potential use in building a climate-neutral society. \par

MC hydrogen liquefaction holds great promise, but its successful implementation faces challenges. If hydrogen gas is precooled by liquid nitrogen, MC hydrogen liquefaction requires MC materials with working temperatures covering the temperature range from the nitrogen condensation point (77 K) to the hydrogen condensation point (20 K) \cite{Park.2015,Park.2017,Matsumoto.2009,Numazawa.2014}. For large-scale applications, cost-effective MC materials with large magnetic entropy and adiabatic temperature changes ($\Delta S_T$  \& $\Delta T_{ad}$) in this temperature range in affordable magnetic fields such as 2 T (generated by Nd-Fe-B permanent magnets \cite{Gutfleisch.2011}) or 5 T (generated by commercial superconducting magnets \cite{Matsumoto.2009}) are required. \par

Rare-earth-based MC intermetallic compounds are competitive candidates for hydrogen liquefaction due to their excellent $\Delta S_T$  and $\Delta T_{ad}$ \cite{Franco.2018,bykov_magnetocaloric_2024}. Rare-earth elements can be classified into two groups: (1) the light rare earths La, Ce, Pr, Nd, and Sm; (2) the heavy rare earths Gd, Tb, Dy, Ho, Er, Tm, Yb, and Lu \cite{Zepf2013}. Whether Eu belongs to light rare-earth elements or not is being debated and is sometimes classified into medium rare-earth elements \cite{Zepf2013}. Furthermore, although Sc and Y do not belong to the lanthanides, they are classified as heavy rare earths due to their chemical and physical similarities \cite{Zepf2013}. 

In terms of cryogenic MC materials, light rare-earth-based intermetallic compounds are often overlooked because they generally show a less impressive MCE than their heavy rare-earth counterparts due to the smaller magnetic moments of the light rare-earth ions \cite{liu2023designing}. However, heavy rare-earth elements such as Ho and Er are highly critical \cite{Coey.2019}, raising questions on the feasibility of using these compounds for large-scale hydrogen liquefaction applications \cite{liu2023designing}. On the contrary, due to the low resource criticality of light rare-earth elements, their intermetallic compounds show great potential to scale up MC hydrogen liquefaction technology \cite{liu2023designing}. A dilemma appears when choosing suitable MC materials for hydrogen liquefaction: materials exhibiting higher performance or containing fewer critical elements? \par

This review focuses on the performance and criticality of rare-earth-based MC intermetallic compounds. In the first part, the influence of the cryogenic temperature on the MC performance of the rare-earth-based intermetallic compounds is explored. In the second part, we focus on tailoring the MC performance of rare-earth-based intermetallic compounds by tuning their Curie temperatures $T_C$. The last part focuses on resolving the dilemma of performance and criticality.  \par

\section{MCE at cryogenic temperatures}\label{MCEatcryogenic}

\subsection{"Giant" second-order MCEs}

\begin{figure*}[ht]
\centering
\includegraphics[width=0.7\linewidth]{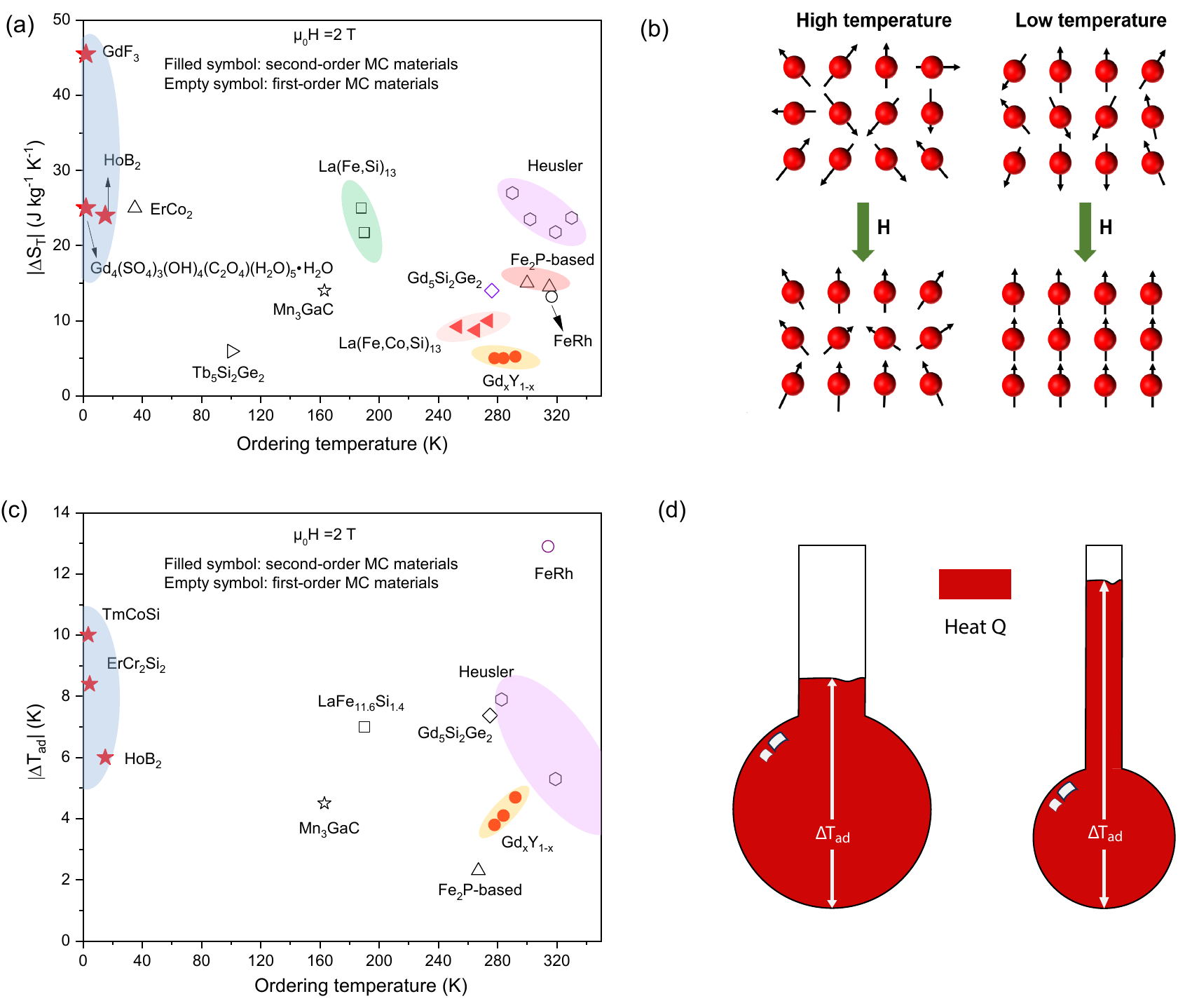}
\caption{ (a) (c) $\Delta S_T$ and $\Delta T_{ad}$ of selected first-order MC materials in comparison with selected second-order MC materials. Data are taken from \cite{Pecharsky.1997b,huang_preparation_2002,Taubel.2020,Taubel.2018,chirkova_giant_2016,Tegus.2002,dung_first-order_2011,Gottschall.2019,saito_magnetocaloric_2007,Balli.2011,hu_direct_2003,Scheibel.2015,Castro.2020,balli2007optimization,chen2015brilliant,han_large_2014,skokov_influence_2013,Fries.2017,Nikitin.1990,Xu.2020,Li.2012b} (seen in the supplementary). (b)(d) Illustrations on how cryogenic temperature influences $\Delta S_T$ and $\Delta T_{ad}$.}\label{Fig2}
\end{figure*}

"Giant" is often a label attached to first-order MC materials such as  \ce{Gd5(Si, Ge)4} alloys \cite{Pecharsky.1997b}, \ce{La(Fe,Si)13}  alloys \cite{Zhang.2000,Liu.2011b,scheibel_hysteresis_2018,shao_high-performance_2017, waske_asymmetric_2015}, Ni-Mn-based Heusler alloys \cite{han2006large,Liu.2012,Krenke.2005,Wei.2015,Wei.2016,Taubel.2020}, \ce{Fe2P}-based alloys \cite{Tegus.2002}, and \ce{Fe-Rh} \cite{Nikitin.1990, chirkova_magnetocaloric_2021,chirkova_giant_2016}, as they show a distinctly large $\Delta S_T$ in relatively low magnetic fields  (e.g. \qty{2}{\tesla}). In comparison, second-order MC materials are often described as "MC materials that exhibit weaker MCEs" \cite{Liu.2022}, although many "giant" MCEs in first-order MC materials, such as the Ni-Mn-based Heusler alloys, are absent under cycling in moderate magnetic fields because of the significant thermal hysteresis associated with first-order phase transition. Second-order MC materials have a fully reversible MCE due to their zero thermal hysteresis. \par

However, some paramagnetic salts show extremely large $\Delta S_T$ even in low magnetic fields \cite{chen2015brilliant}. \textbf{\Cref{Fig2}} (a) compares the $\Delta S_T$ (without considering cyclic performance) of the selected first-order MC materials with the selected second-order MC materials, including the rare-earth-based \ce{GdF3} \cite{chen2015brilliant}, \ce{Gd4(SO4)3(OH)4(C2O4)(H2O)5*H2O} \cite{han_large_2014}, and \ce{HoB2} \cite{Castro.2020} with ordering temperatures below \qty{20}{K}. It is true that the near-room-temperature second-order MC materials \ce{Gd_{x}Y_{1-x}} \cite{Gottschall.2019} (magnetic entropy change in magnetic fields of 0.8 T can be found in reference \cite{kito2007study}) and \ce{La(Fe,Co,Si)13} \cite{balli2007optimization} show the smallest $\Delta S_T$ in the plot. However, the second-order MC material \ce{GdF3} exhibits the largest $\Delta S_T$ of \qty{45.5}{\joule\per\kelvin\per\kilogram} at 2 K in \textbf{\Cref{Fig2}} (a). \ce{Gd4(SO4)3(OH)4(C2O4)(H2O)5*H2O} and \ce{HoB2} that exhibit second-order phase transitions show comparable $\Delta S_T$ to \ce{ErCo2}, a first-order MC material that is described as one of the best MC materials for hydrogen liquefaction \cite{Singh.2007}.\par

These observations have broken the stereotype that "second-order MC materials show smaller $\Delta S_T$ than the giant first-order MC materials". The cryogenic temperatures have a significant influence on $\Delta S_T$. \textbf{\Cref{Fig2}} (b) is a schematic diagram illustrating how cryogenic temperature influences $\Delta S_T$. 
The lower the temperature, the weaker the thermal motion, and the magnetic moments are more easily aligned by external magnetic fields, resulting in a larger $\Delta S_T$. \par

As an equally important MC parameter \cite{Gottschall.2019}, $\Delta T_{ad}$ is also greatly influenced by the cryogenic temperature. \textbf{\Cref{Fig2}} (c) compares the $\Delta T_{ad}$ of the second-order MC materials \ce{TmCoSi} \cite{Xu.2020}, \ce{ErCr2Si2} \cite{Li.2012b}, and \ce{HoB2} \cite{Castro.2020} with the selected first-order and second-order MC materials at elevated temperatures. In contrast to $\Delta S_T$ (\textbf{\Cref{Fig2}} (a)) where the largest value occurs at low temperature, the near-room-temperature first-order MC material \ce{FeRh} shows the largest value of $\Delta T_{ad}$ \cite{Nikitin.1990}. However, the second-order MC materials ordering at low temperatures (\ce{TmCoSi}, \ce{ErCr2Si2}, and \ce{HoB2}) exhibit a $\Delta T_{ad}$ that is comparable to or greater than all the giant first-order MC materials excluding \ce{FeRh} in \textbf{\Cref{Fig2}} (c), while room-temperature second-order MC materials \ce{Gd_{x}Y_{1-x}} shows a smaller $\Delta T_{ad}$ than many first-order MC materials. \textbf{\Cref{Fig2}} (d) illustrates how the heat capacity influences $\Delta T_{ad}$. Since the heat capacity $C_p$ at low temperatures is smaller than at room temperature and the heat $Q = C_p \, dT$, a smaller $C_p$ would result in a larger temperature change $dT$. \par

The above discussion indicates a significant influence of the cryogenic temperature on the MCE as second-order MC materials can also show a strong effect at low temperatures as the giant first-order ones. \par

\subsection{Increasing trends of MCEs}\label{increasingtrend}

\begin{figure}[ht]
	\centering
    \includegraphics[width =0.75\linewidth]{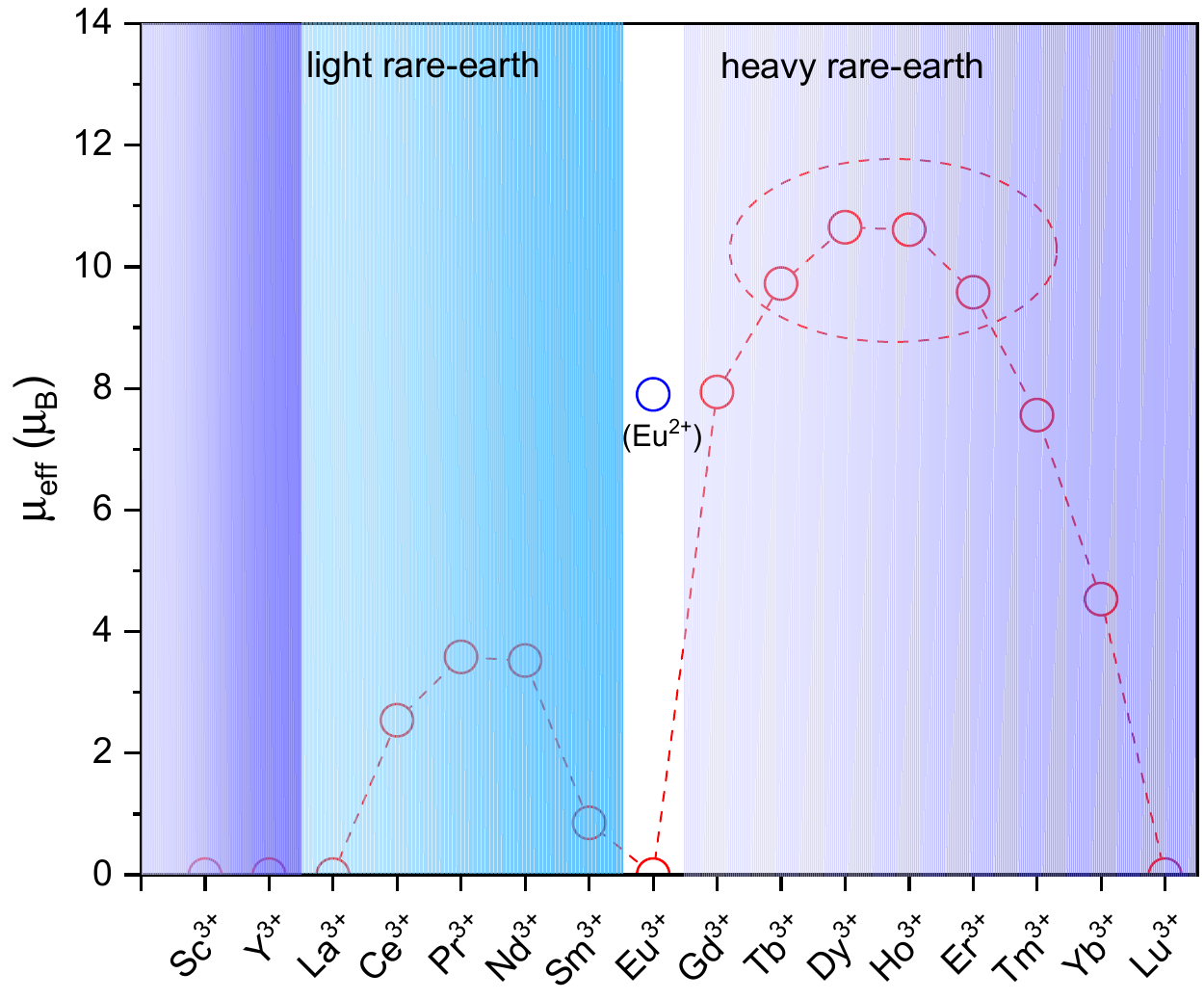}
    \caption{Effective magnetic moments of rare-earth ions from Hund's rule. Data are taken from reference \cite{coey2010magnetism}.}
    \label{Fig3}
\end{figure}

This subsection focuses on describing how cryogenic temperature influences MCEs by studying the correlations of $\Delta S_T$ and $\Delta T_{ad}$ with respect to $T_C$. Ideally, a material system that only varies in $T_C$ with the other physical parameters remaining constant is needed for this purpose, since, in addition to $T_C$, $\Delta S_T$ and $\Delta T_{ad}$ are influenced by many extrinsic factors such as secondary phases and texture that belong to the microstructural features\cite{Gutfleisch.2016,Zhang.2015,Zhang.2020, doring_diffusion_2022}, and many intrinsic factors such as magnetic moment \cite{Liu.2022,liu2023designing} and the nature of phase transition \cite{Balli.2017,Kitanovski.2020,Gutfleisch.2011,smith2012materials}. The chemical and physical similarities of the rare-earth elements make this study possible. \textbf{\Cref{Fig3}} plots the effective magnetic moments of all the lanthanide rare-earth ions. Heavy rare-earth ions \ce{Gd^3+}, \ce{Tb^3+}, \ce{Dy^3+}, \ce{Ho^3+}, \ce{Er^3+}, and \ce{Tm^3+} possess effective magnetic moments larger than 7 $\mu_B$. In particular, the effective magnetic moments of \ce{Tb^3+}, \ce{Dy^3+}, \ce{Ho^3+}, and \ce{Er^3+} are rather close, ranging around 10 $\mu_B$, as indicated by the ellipse.  \par

\begin{figure*}[hp]
	\centering
    \includegraphics[width = \linewidth]{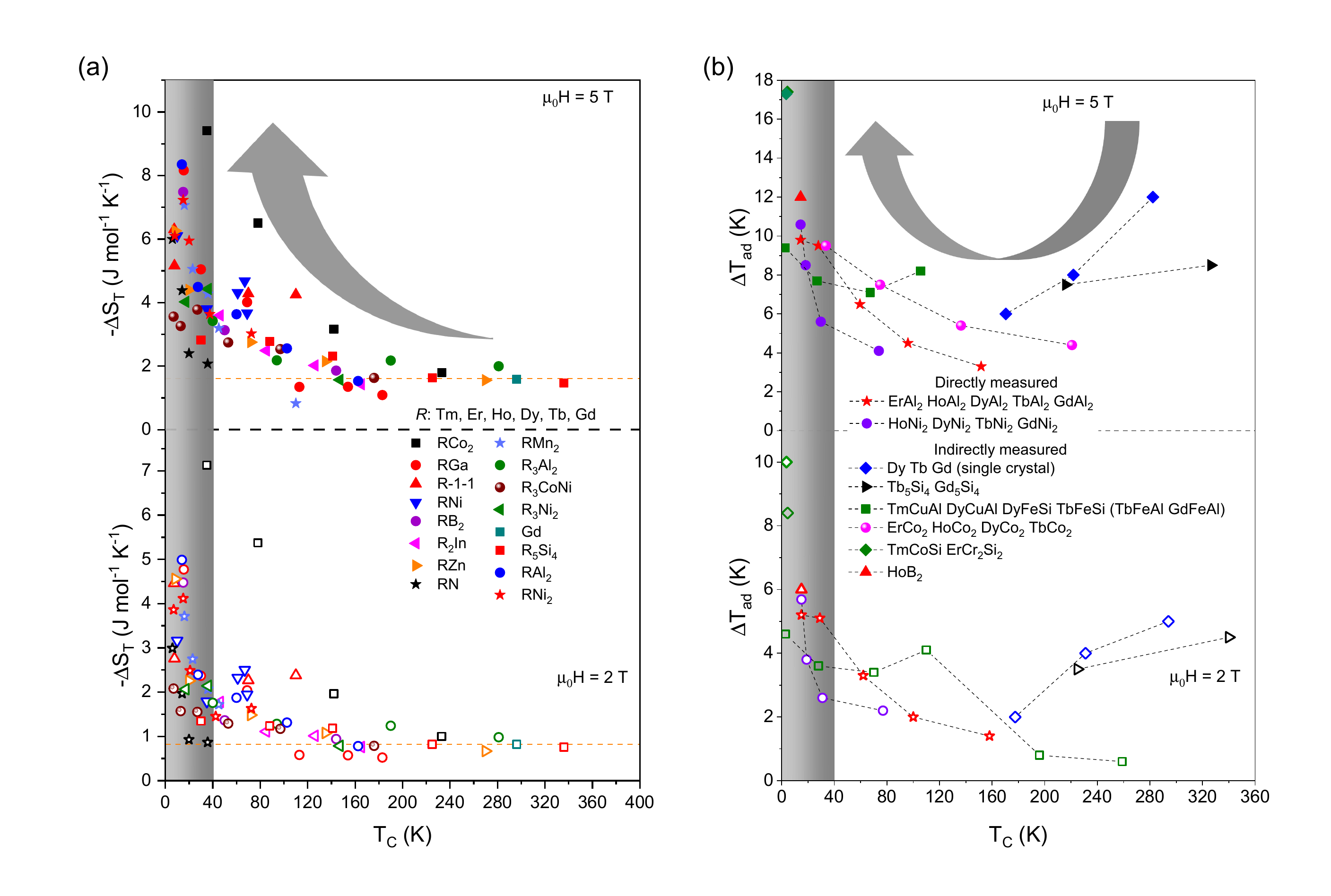}
    \caption{(a) (b) Literature review on the maximum $\Delta S_T$ and $\Delta T_{ad}$ of heavy rare-earth-based MC intermetallic compounds in magnetic fields of 2 (top) and 5 T (bottom). The unit of $\Delta S_T$ is \qty{}{\joule\per\mole\per\kelvin} (per one mole of rare-earth atoms). The orange dashed lines highlight the $\Delta S_T$ of the benchmark MC material Gd. Data are taken from \cite{Gottschall.2019,Zeng.2016, singh_measurement_2007,Balli.2007,Singh.2005,Tohei.2004,Zheng.2014,Zheng.2012,Chen.2010,Chen.2009, Mo.2013,Zhang.2013,Dong.2009,Mo.2013b,Rajivgandhi.2017,Rajivgandhi.2016,Kurian.2020,Han.2010,Meng.2012,Zhang.2009,Zhang.2009b,Zhang.2009c,Zhang.2011,Pecharsky.2002,Wang.2017,Li.2015,Li.2017,Li.2015b,Shinde.2015,Nakagawa.2006,Marcos.2004,Zuo.2013,Pecharsky.2002b,Zhang.2014,Li.2015c,Zhang.2012,Provino.2016,Herrero.2021,Dong.2011,spichkin_preparation_2001,Ivtchenko.2000,Singh.2010,Arnold.2009,Balli.2011,Zhou.2006,Duc.2002,Liu.2022,Kuzmin.1993,Nobrega.2006,Singh.2007Tm,Wada.2001,Xu.2020,Li.2012b,Kastil.2014} (seen in the supplementary)}.\label{Fig4}
\end{figure*}

Despite the similar effective magnetic moments of \ce{Tb^3+}, \ce{Dy^3+}, \ce{Ho^3+}, and \ce{Er^3+}, their alloys generally exhibit significantly different values of $\Delta S_T$ and $\Delta T_{ad}$. \textbf{\Cref{Fig4}} (a) shows the review of $\Delta S_T$ (in units of \qty{}{\joule\per\mole\per\kelvin}) with respect to $T_C$ for heavy rare-earth-based MC intermetallic compounds in magnetic fields of 2 and 5 T. As indicated by the arrow, the values of $\Delta S_T$ exhibit an increasing trend with decreasing $T_C$. This trend becomes more pronounced in the vicinity of the hydrogen condensation point, with giant values of $\Delta S_T$ observed near \qty{20}{K}. In addition, many heavy rare-earth-based MC intermetallic compounds with a cryogenic $T_C$ are observed to exhibit a larger maximum $\Delta S_T$ than \ce{Gd}, the benchmark material used for room-temperature refrigeration. \par

\textbf{\Cref{Fig4}} (b) shows the review of $\Delta T_{ad}$ with respect to $T_C$ for heavy rare earth-based MC intermetallic compounds in magnetic fields of 2 and 5 T.  As indicated by the arrow, $\Delta T_{ad}$ increases with decreasing $T_C$ in the cryogenic temperature range, exhibiting the same trend as $\Delta S_T$. However, from room temperature to approximately \qty{150}{K}, the increasing trend is absent. On the contrary, a decreasing trend is observable. It should be emphasized that the data in this range are not abundant and more measurements on $\Delta T_{ad}$ need to be made in the future to confirm this decreasing trend.  \par

The different behaviour of $\Delta S_T$ and $\Delta T_{ad}$ with respect to $T_C$ can be understood by the approximation equation:
\begin{equation}
    \Delta T_{ad}(T_C) \approx -\frac{T_C \, \Delta S_T(T_C)}{C_p} .
\end{equation}
As $T_C$ decreases, $\Delta S_T (T_C)$ increases but $C_p$ decreases. Near room temperature, most materials follow the Dulong-Petit law and their lattice heat capacities stay almost constant, resulting in a decreasing $\Delta T_{ad}$, while towards low temperatures, the heat capacities of many materials undergo a sharp decrease, resulting in an increasing $\Delta T_{ad}$. \par

\subsection{Mean-field approach and power laws}

The large values of $\Delta S_T$ and $\Delta T_{ad}$ of second-order MC materials at low temperatures, and the increasing trends of $\Delta S_T$ and $\Delta T_{ad}$ with decreasing $T_C$, can be well explained by mean-field theory. \par

The total entropy $S(T, H)$ of a MC material can be expressed as:
\begin{equation}
    S(T, H) = S_m(T, H) + S_l(T)+S_e(T) \, ,
\end{equation}
where $S_m(T, H)$ is the magnetic entropy, $S_l(T)$ and $S_e(T)$ are lattice and electronic entropies and are assumed to be independent on external magnetic field $H$.

 $\Delta S_T$ and $\Delta T_{ad}$ can be obtained from the total entropy curve $S(T, H)$ by equation \cite{Pecharsky.1999}:
 \begin{equation}\label{dSanddT}
 \begin{split}
        &\Delta S_T (T, H) = S(T,H) - S(T,0)\, , \\
    	&\Delta T_{ad} (T,H) = T(S,H) - T(S,0) \, , 
 \end{split}
\end{equation}
where $T(S, H)$ is the inverse function of $S(T, H)$. \par

Magnetic entropy $S_m$ can be obtained from mean-field theory. The partition function of a canonical ensemble is
\begin{equation}
    Z = \sum_n \exp{\left( -\frac{E_n}{k_B T} \right)} \, ,
\end{equation}
where $E_n$ is the eigenvalue of the Hamiltonian $\hat{H}$ of the system, and $k_B$ is the Boltzmann constant. Considering a system with one paramagnetic atom, $\hat{H}$ takes the form:
\begin{equation}
    \hat{H} = -\hat{M_J} \mu_0 \Vec{H} \, ,
\end{equation}
where $\Vec{H}$ is the vector of the magnetic field, $\mu_0$ is the vacuum permeability and $\hat{M}$ is the atom magnetic moment operator given by:
\begin{equation}
    \hat{M_J} = g_J \mu_B \hat{J} \, ,
\end{equation}
where $g_J$ being the Landé g-factor, $\mu_B$ the Bohr magneton, and $\hat{J}$ the total angular momentum operator. The magnetic moment of an atom is
\begin{equation}
    M_J = g_J \mu_B J \, ,
\end{equation}
where  $J$ is the total magnetic momentum number. The projection of $J$ along the magnetic field is $m$, which takes the values $J, \, J-1, \, ... , \, -J$. Thefore, the eigenvalue of $\hat{H}$ is
\begin{equation}
    E_m = -g_J \mu_B m \mu_0 H \, .
\end{equation}
Let
\begin{equation}
    y = \frac{M_J\mu_0 H}{k_B T} \, ,
\end{equation}
then
\begin{equation}
    \begin{split}
    Z(x) &= \sum_{m=-J}^{J} \exp{\left( \frac{my}{J} \right)} \\
         &= \frac{\sinh{\left ( \frac{2J+1}{2J} y\right )}}{\sinh{\left ( \frac{1}{2J} y\right )}} \, ,
    \end{split}
\end{equation}
For a system with $N$ atoms, the partition function is $Z_N(y) = (Z(y))^N$. The free energy $F$ of a canonical system is given by
\begin{equation}
    F= -N k_B T \ln{(Z(y))}
\end{equation}
The magnetic entropy $S_m$ is given by \cite{Tishin.2003}:
\begin{equation}\label{eqSm}
    \begin{split} 
    S_m &= -\frac{\partial F}{\partial T}  \\
    &= N_M k_B \left [ \ln{ \left ( \frac{\sinh{\left ( \frac{2J+1}{2J}y \right )} }{\sinh{\left ( \frac{1}{2J}y \right )} }\right ) - yB_J(y)}  \right ] ,
    \end{split}
\end{equation}
with $N_M$ the number of “magnetic atoms”, and $B_J (y)$ the Brillouin function.

The above calculation considers a canonical system with paramgnetic atoms. For a system with ferromagnetic atoms, the internal contribution of $n_wM$ ($n_w$ is the Weiss coefficient and $M$ is the magnetization) to the magnetic field. $M$ can be expressed as:
\begin{equation}
    M = N M_J B_J(y) \, .
\end{equation}
The Weiss coefficient $n_w$ is given by
\begin{equation}
    n_w = \frac{3 k_B T_C}{N \mu_0 g_J^2 \mu_B^2 J(J+1)} \, .
\end{equation}
Knowing $n_wH$, $y$ is given by
\begin{equation}
    \begin{split}         
    y &= \frac{M_J \mu_0 (H + n_w M)}{k_B T} \\
      &= \frac{g_J J \mu_B \mu_0 H + \frac{3J}{J+1}k_B T_C B_J(y)}{k_BT} \, .
    \end{split}
\end{equation}
The process to derive \textbf{\Cref{eqSm}} can also be found in \cite{Tishin.2003,de_oliveira_theoretical_2010}.

The equation for calculating the lattice entropy $S_l$ is given as \cite{Tishin.2003}:
  \begin{equation}\label{lattice}
  \begin{split}
      S_l = & -3Nk_B \left[ \ln{ \left ( 1 - \exp{\left ( -\frac{T_D}{T} \right )} \right )}  \right] \\
      &+12Nk_B \left( \frac{T}{T_D}\right)^3 \int_0^{T_D/T} \frac{x^3}{\exp{(x)}-1} dx \, ,
  \end{split}
  \end{equation}
  where $T_D$ is the Debye temperature, $N$ is the total number of atoms, and $x$ is a variable taking values from $0$ to $T_D/T$. \par
  
The electronic entropy is give as 
\begin{equation}
    S_e = \gamma \, T \,
\end{equation}
where $\gamma$ is the Sommerfeld coefficient. The electronic entropy $S_e$ is usually small compared to $S_m$ and $S_l$ and only plays an important role at sufficiently low temperatures \cite{kittel_introduction_2018}. 

\begin{figure*}[hb]
	\centering
    \includegraphics[width = \linewidth]{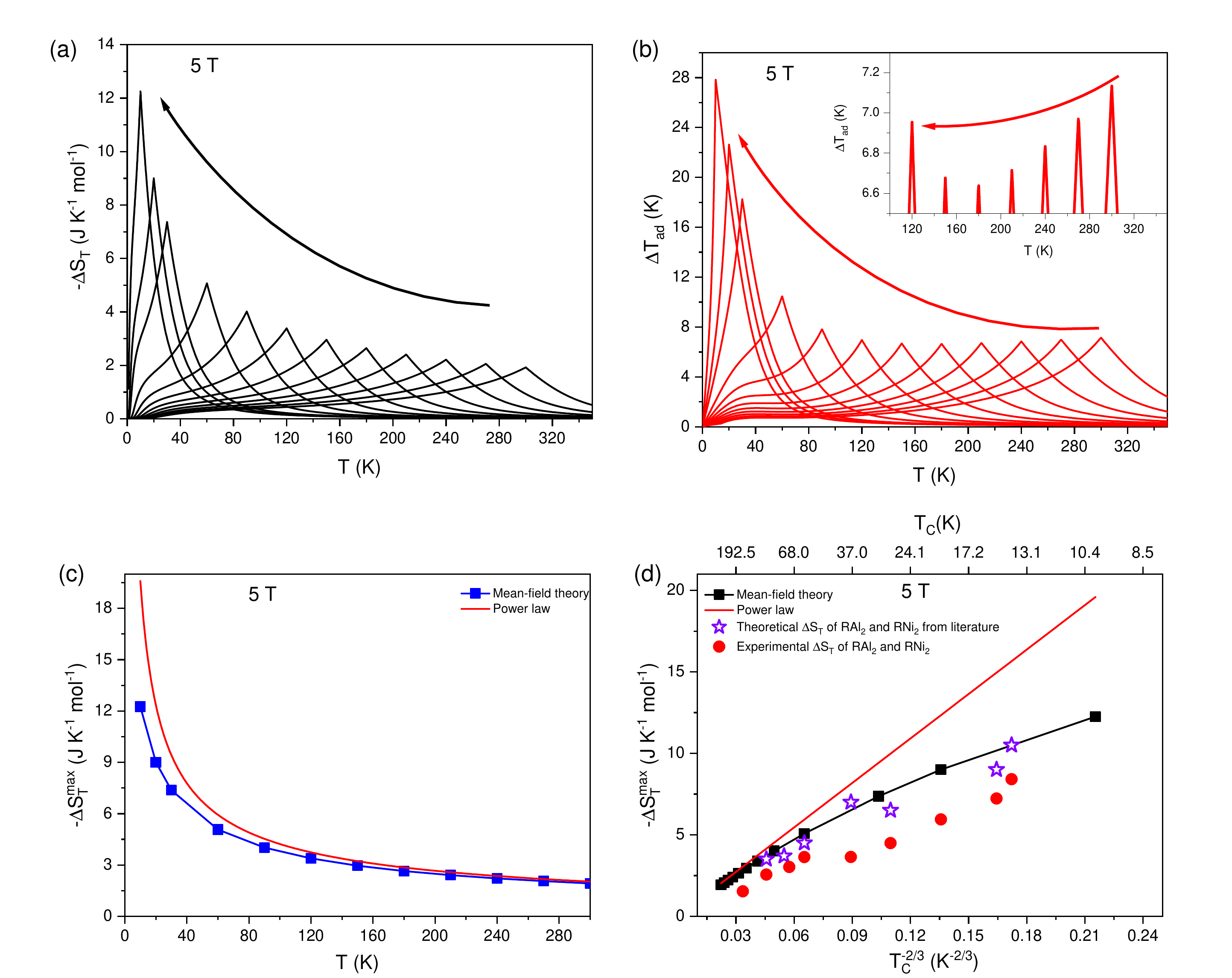}
    \caption{(a) (b) $\Delta S_T$ and $\Delta T_{ad}$ from the mean-field approach in magnetic fields of 5 T by taking $g_J$, $J$, and $T_D$ as 4/3, 7.5 and \qty{384}{\kelvin}, respectively. (c) (d) Comparison of $\Delta S_T$ calculated directly from the mean-field approach with $\Delta S_T$ given by the power law of $\Delta S_T \propto T_C ^{-2/3}$.  The top axis in (d) shows the corresponding values of $T_C$. Furthermore, the experimental and theoretical $\Delta S_T$ of \ce{RAl2} and \ce{RNi2} (R= Gd, Tb, Dy, Ho, and Er) are plotted in (d). Data are taken from references \cite{von_ranke_influence_2001-1,von_ranke_influence_2001,Liu.2022} (seen in the supplementary). }
    \label{Fig5}
\end{figure*}

 \textbf{\Cref{Fig5}} (a) and (b) plot the maximum $\Delta S_T$ and $\Delta T_{ad}$ calculated from the mean-field approach by taking $N_M$ to be 1/3 \unit{\mole}, and the magnetic parameters to be the same as \ce{Dy^3+} to make the calculation correspond to \ce{DyAl2}. Consistent with the literature review presented in \textbf{\Cref{Fig4}}, the mean-field approach provides theoretical evidence for the observations that the giant values of $\Delta S_T$ and $\Delta T_{ad}$ exist at low temperatures and that $\Delta S_T$ and $\Delta T_{ad}$ increase as $T_C$ decreases in the cryogenic temperature range. In addition, the decreasing trend of $\Delta T_{ad}$ with decreasing $T_C$ near room temperature where the Dulong-Petit law applies is validated. It should be mentioned that Debye temperature $T_D$ is also a factor that could have a significant influence on $\Delta T_{ad}$, as described in \cite{liu_role_2023}.\par

The feature that second-order MC materials can show "giant" values of $\Delta S_T$ at low temperatures and the increasing trend of $\Delta S_T$ with the decreasing $T_C$ can also be well understood by the power laws of  $\Delta S_T$  with respect to $T_C$. Three power laws of  $\Delta S_T$ and $\Delta T_{ad}$  with respect to $T_C$ have been summarized and validated \cite{belo_curie_2012,lyubina_magnetic_2011,kuzmin_magnetic_2011,Gottschall.2019.Gd}:
\begin{align}
     & \Delta S_T(H) \propto H^{\frac{2}{3}}  \label{linear1}\, ,  \\
     & \Delta S_T(T_C) \propto T_C ^{-\frac{2}{3}} \label{linear2} \, ,  \\
     & \Delta T_{ad}(H) \propto H^{\frac{2}{3}} \, . 
\end{align}

The complete form of \textbf{\Cref{linear1}} and \textbf{\Cref{linear2}} is given by Oesterreicher \textit{et al.} as \cite{Oesterreicher.1984}:
\begin{equation}\label{approx}
\begin{split}
    \Delta S_T(T_C, H) =& -\frac{3}{2}R\,J(J+1)\\ 
     &\left ( \frac{10}{9}\frac{g_J\mu_0\mu_BH}{[(J+1)^2 + J^2] k_B T_C} \right )^{\frac{2}{3}} \, .
\end{split}
\end{equation}
Mathematically, the function of $f(x) = x^{-\frac{2}{3}} $ is a power function that increases with the  decreasing $x$, and the increasing trend gets stronger towards lower $x$.

 \textbf{\Cref{Fig5}} (c) and (d) compare the directly calculated $\Delta S_T$ from \textbf{\Cref{eqSm}} and the approximated $\Delta S_T$ from the power law (\textbf{\Cref{approx}}) under a magnetic field change of 5 T. From room temperature down to about 120 K, both calculations fit well. However, below 120 K, the $\Delta S_T$ approximated from the power law starts to deviate from the directly calculated $\Delta S_T$ from the mean-field approach. Below 20 K, the power law predicts significantly larger values of $\Delta S_T$, indicating that the power law of $\Delta S_T \propto T_C^{-2/3}$ is a good approximation near room temperature, but not accurate enough below 20 K. \par
 
 In addition, theoretical and experimental $\Delta S_T$ of \ce{RAl2} and \ce{RNi2} (R = Gd, Tb, Dy, Ho, and Er) are plotted in \textbf{\Cref{Fig5}} (d). The theoretical $\Delta S_T$ of \ce{RAl2} and \ce{RNi2} are taken from the references \cite{von_ranke_influence_2001-1,von_ranke_influence_2001}, with the crystal electrical field taken into account. The experimental data for the \ce{RAl2} and \ce{RNi2} polycrystal samples are taken from the reference \cite{Liu.2022}. The theoretical $\Delta S_T$ from the literature shows a similar value to that from the mean-field approach in this work and roughly follows the $T_C^{-2/3}$ power law. The experimental data show a lower value but still roughly follow a linear path. \par
 
From the above discussion, it can be seen that the cryogenic temperature is beneficial to MCEs: the maximum $\Delta S_T$ and $\Delta T_{ad}$ show increasing trends with decreasing $T_C$ in the cryogenic temperature range, resulting in distinctly large values in the vicinity of the hydrogen condensation point. Because of this advantage at cryogenic temperatures, many second-order rare-earth-based MC materials are competitive candidates for MC hydrogen liquefaction. \par

\pagebreak
\section{Tailoring MCE}

 \begin{figure*}[hp]
	\centering
    \includegraphics[width = 0.95\linewidth]{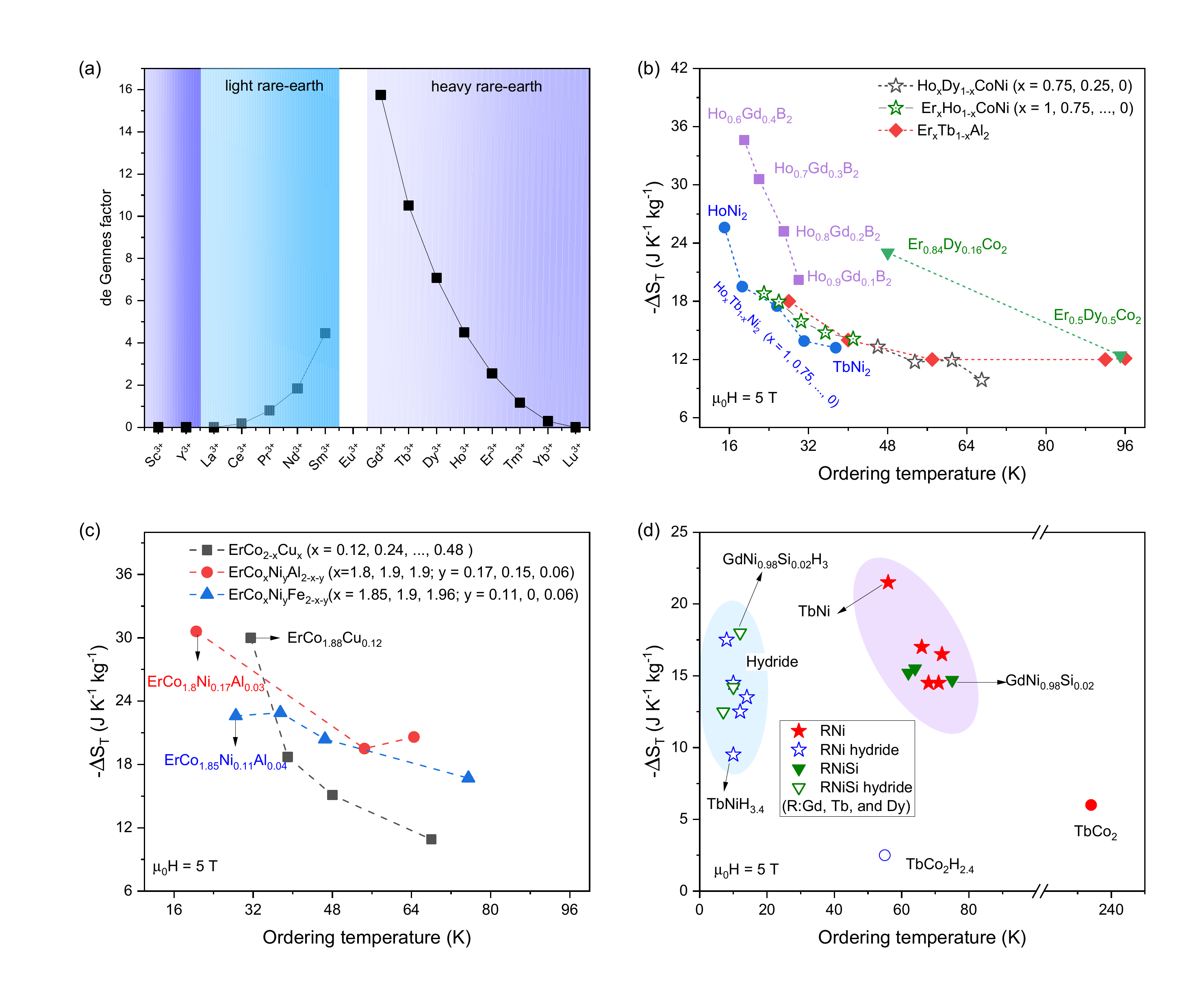}
    \caption{(a) De Gennes factors of rare-earth ions. Values are taken from \cite{coey2010magnetism}. (b) (c) (d) $\Delta S_T$ with respect to the ordering temperatures in the cases of mixing rare-earth elements, doping the non-rare-earth site, and hydrogenation (the light red shadow marks the compounds before hydrogenation and the light blue shadow marks the corresponding hydrides). Data are taken from \cite{Khan.2010,Cwik.2021,Zhang.2023,Zhu.2011,Castro.2021.Gd,Lu.2022,Tang.2022,Tereshina.2023,Politova.2022,Chzhan.2020,Chzhan.2021,Lushnikov.2018,Smarzhevskaya.2014,Smarzhevskaya.2015} (seen in supplementary).}
    \label{Fig6}
\end{figure*}

The MC hydrogen liquefaction requires a series of MC materials with working temperatures covering the range from 77 to 20 K \cite{Liu.2022}. In magnetic fields of 2 and 5 T, almost no single MC material can cover such a large temperature range with sufficient $\Delta S_T$ and $\Delta T_{ad}$ \cite{Liu.2022}. It is necessary to design a sequence of materials with their MCEs adapted to cover the full temperature range required by MC hydrogen liquefaction \cite{Liu.2022,liu2023designing}.  \par

The above subsections demonstrate that the maximum $\Delta S_T$ and $\Delta T_{ad}$ are highly correlated with $T_C$, suggesting that MCEs can be tuned to fulfill the requirements of MC hydrogen liquefaction by tailoring the $T_C$. One of the advantages of rare-earth-based MC intermetallic compounds is their easily adjustable ordering temperatures \cite{Liu.2022,liu2023designing}. Tuning the $T_C$ of rare-earth-based MC intermetallic compounds by mixing different rare-earth elements at the rare-earth sublattice is the most common way due to the chemical and physical similarities of rare-earth elements that benefit the mixing method \cite{Liu.2022,liu2023designing}. \par

The physical mechanism of this method is that ordering temperatures of rare-earth-based intermetallic compounds are highly correlated with the de Gennes factor \cite{coey2010magnetism}. As a result of spin-orbital coupling, $\Vec{J}$ is the good quantum number instead of $\Vec{S}$. Therefore, $\Vec{S}$ needs to be projected onto the total angular momentum vector $\Vec{J}$ \cite{coey2010magnetism}:
\begin{align}
    &\Vec{L} + 2 \Vec{S} = g_J \, \Vec{J} \, ,\\
    &\Vec{J} = \Vec{L} + \Vec{S} \, , \\
    &\Vec{S} = (g_J -1) \Vec{J}.
\end{align}
$\Vec{L}$ is the orbital angular momentum vector and $g_J$ is the Landé g-factor. The de Gennes factor $G$ is defined as \cite{coey2010magnetism}:
\begin{equation}
    \begin{split}
        G &= (g_J -1)\Vec{J}^2 \\
        & = (g_J -1)J(J+1) \, .
    \end{split}
\end{equation} \par

The Curie temperature is connected with de Gennes factor $G$ via equation \cite{coey2010magnetism}:
\begin{equation}\label{deCurie}
    T_C = \frac{2Z\mathcal{J}G}{3k_B}.
\end{equation}
$\mathcal{J}$ is the Heisenberg exchange constant and $Z$ is the number of the nearest atoms. When the band structure and lattice constant stay the same, $T_C$ usually scales with the de Gennes factor.

\textbf{\Cref{Fig6}} (a) shows the de Gennes factor of rare-earth ions \cite{coey2010magnetism}. For heavy rare-earth ions, Gd exhibits the largest de Gennes factor $G$, and as the atomic number increases,  $G$ decreases gradually. For \ce{Lu^3+}, $G$ becomes zero. For the light rare-earth ions, the trend is inverted and $G$ increases gradually with increasing atomic number. \textbf{\Cref{Fig6}} (b) displays the examples of mixing different rare-earth elements to tune the $T_C$ of the rare-earth-based intermetallic compounds for tailoring the MCE to cover the temperature range of 77 $\sim$ 20 K. Agreeing with the conclusion in \textbf{\Cref{increasingtrend}}, all the material systems exhibit an increasing trend of $\Delta S_T$ upon tuning the $T_C$ to lower temperatures. \par

 \begin{figure*}[t]
	\centering
    \includegraphics[width = 0.9\linewidth]{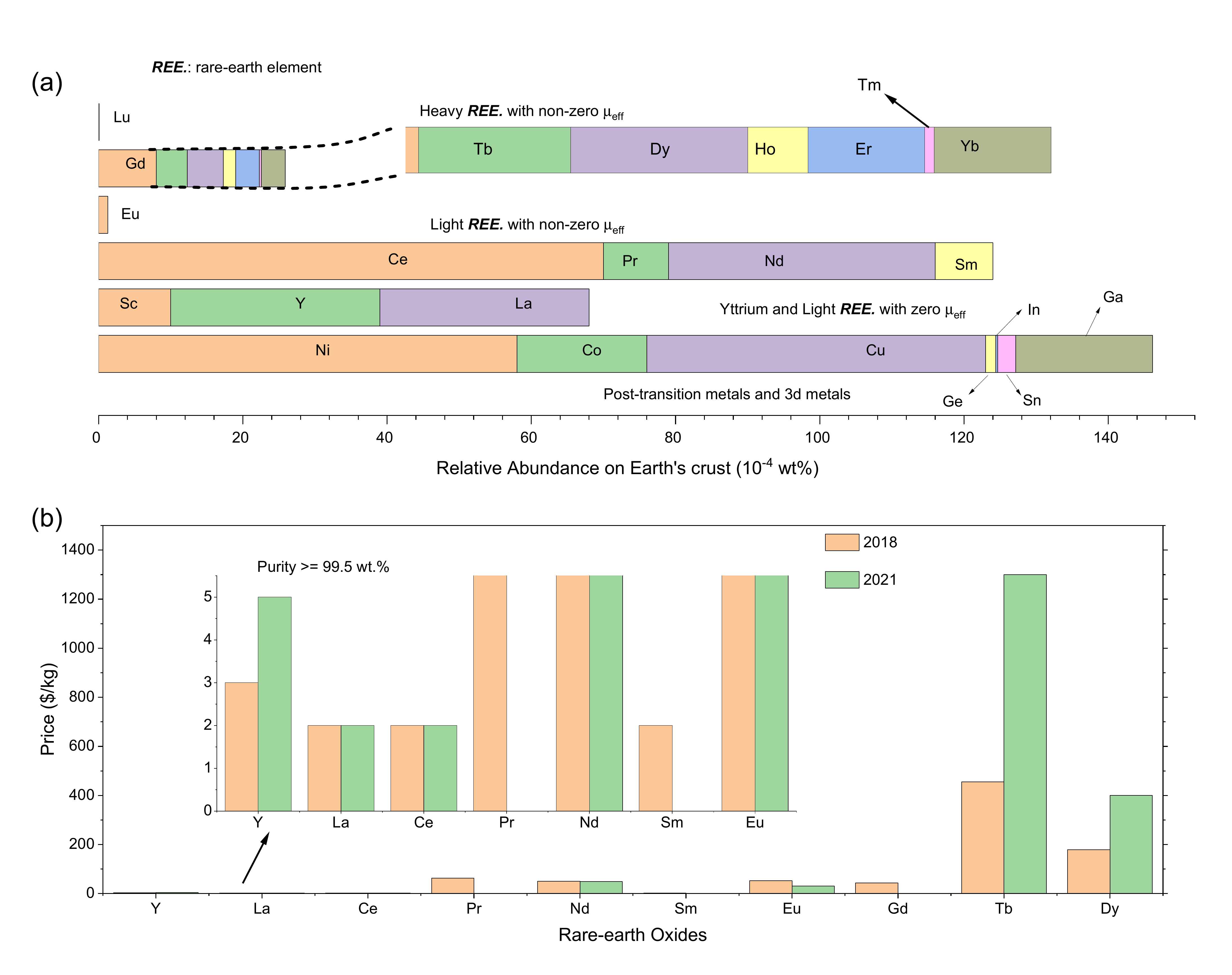}
    \caption{(a) Comparing the relative abundances of rare-earth elements and 3d metals (Ni, Co, and Cu) on earth's crust. The rare-earth elements are categorized into non-heavy rare-earth elements with zero magnetic moment (Sc, Y, and La), heavy rare-earth elements with non-zero magnetic moments (Gd, Tb, Dy, Ho, Er, Tm, and Yb), light rare-earth elements with non-zero magnetic moments (Ce, Pr, Nd, and Sm), Eu, and Lu. Data are taken from \cite{yaroshevsky_abundances_2006}.  (b) Prices of some of the rare-earth oxides in years 2018 and 2021. Data are taken from \cite{noauthor_critical_2022, noauthor_mineral_2022}.}
    \label{Fig7}
\end{figure*}

Except for adjusting the de Gennes factor to tune the Curie temperature, based on \textbf{\Cref{deCurie}}, $T_C$ can also be tuned by adjusting the exchange constant $\mathcal{J}$. This can be done by doping the non-rare-earth sublattice sites. \textbf{\Cref{Fig6}} (c) shows three examples: (1) \ce{ErCo2_{-x}Cu_{x}} \cite{Lu.2022},(2) \ce{ErCo2_{-x-y}Ni_{x}Al_{y}} \cite{Tang.2022}, and (3) \ce{ErCo2_{-x-y}Ni_{x}Fe_{y}} \cite{Tang.2022}. In the case of \ce{ErCo2_{-x}Cu_{x}}, the ordering temperatures are tuned to approach 77 K by doping Cu. As the ordering temperature increases, the nature of the phase transition changes from first-order ($x \leq 0.24$) to second-order ($x \geq 0.36$ ) \cite{Tang.2022}. In the case of \ce{ErCo2_{-x-y}Ni_{x}Al_{y}} and \ce{ErCo2_{-x-y}Ni_{x}Fe_{y}}, the ordering temperatures are successfully tuned to cover the temperature range of 77 $\sim$ 20 K \cite{Tang.2022}. \par

A third method to tune $T_C$ of the rare-earth-based intermetallic compounds is hydrogenation. \textbf{\Cref{Fig6}} (d) plots $\Delta S_T$ of \ce{TbCo2}, \ce{RNi}, \ce{RNiSi} (R: heavy rare-earth elements), and their corresponding hydrides. The $T_C$ of \ce{TbCo2} is tuned from about 227 K to about 55 K by hydrogenation \cite{Politova.2022}. The $T_C$ of \ce{RNi} and \ce{RNiSi} are tuned by hydrogenation from well above 20 K to below 20 K \cite{Tereshina.2023,Chzhan.2020,Chzhan.2021,Lushnikov.2018,Smarzhevskaya.2014,Smarzhevskaya.2015}. Unlike the \ce{La(Fe,Si)13} compounds that increase their transition temperatures after hydrogenation \cite{fujita2003itinerant, krautz_systematic_2014}, these samples in \textbf{\Cref{Fig6}} (d) decrease their transition temperatures once hydrogenated.\par

In disagreement with the conclusion made in \textbf{\Cref{increasingtrend}}, most rare-earth-based hydrides exhibit a lower $\Delta S_T$ than their base compounds, although the hydrides have a lower $T_C$. Compared to the other two methods, hydrogenation is not as intensively investigated. For improvement of $\Delta S_T$ of the rare-earth-based hydrides, the changes in microstructure, lattice expansion, and exchange constant $\mathcal{J}$ after hydrogenation need to be understood more thoroughly. \par

\section{Criticality as a critical challenge}

In the previous sections, the performance and methods for tailoring the MCE of rare-earth-based intermetallic compounds in the cryogenic temperature range of 77 $\sim$ \qty{20}{K} are discussed mainly by reviewing the heavy rare-earth-based intermetallic compounds. However, heavy rare-earth elements have a high resource criticality, which limits their potential for large-scale applications \cite{Liu.2022}. \textbf{\Cref{Fig7}} (a) shows the relative abundances of the rare-earth elements compared to Ni, Co, and Cu. Light rare-earth metals are much more abundant in the Earth's crust than heavy rare-earth metals. In fact, some rare-earth elements are not "rare" at all: Ce is as abundant as Ni, and the total abundance of Pr and Nd is similar to that of Cu. All of Ce, Nd, Y, and La are significantly more abundant than Co. \par

The prices of the rare-earth oxides can partly reflect the resource criticality of the rare-earth metals. As shown in \textbf{\Cref{Fig7}} (b), terbium and dysprosium oxides are much more expensive than praseodymium and neodymium oxides, and their prices also vary greatly \cite{noauthor_mineral_2022,noauthor_critical_2022}. The oxides of yttrium, cerium, lanthanum, and samarium are much cheaper than the oxides of praseodymium and neodymium \cite{noauthor_critical_2022,noauthor_mineral_2022}. Although europium and gadolinium oxides are as cheap as praseodymium and neodymium oxides, a large increase in price owing to the potentially large demand for MC hydrogen liquefaction should be taken into consideration. \par

It should be emphasized that criticality is much more than just simple geological abundances. Factors such as mining, beneficiation, hazardous by-products, separation (and their social and ecological consequences along this value chain), geopolitics, trade restrictions, and monopolistic supply in terms of demand versus supply, must be understood and quantified in terms of life-cycle-analysis and life-cycle-costing \cite{Gauss2021rare}. A more focused report is in the planning of our group. \par

\subsection{A trade-off between performance and criticality}

Focusing on balancing the performance and criticality, this subsection studies light rare-earth-based second-order MC intermetallic compounds and the "mixed" rare-earth-based (substituting highly critical heavy rare-earth elements with yttrium and light rare-earth elements) intermetallic compounds.  \par

\subsubsection{Light rare-earth-based second-order MC materials}

 \begin{figure*}[ht]
	\centering
    \includegraphics[width = \linewidth]{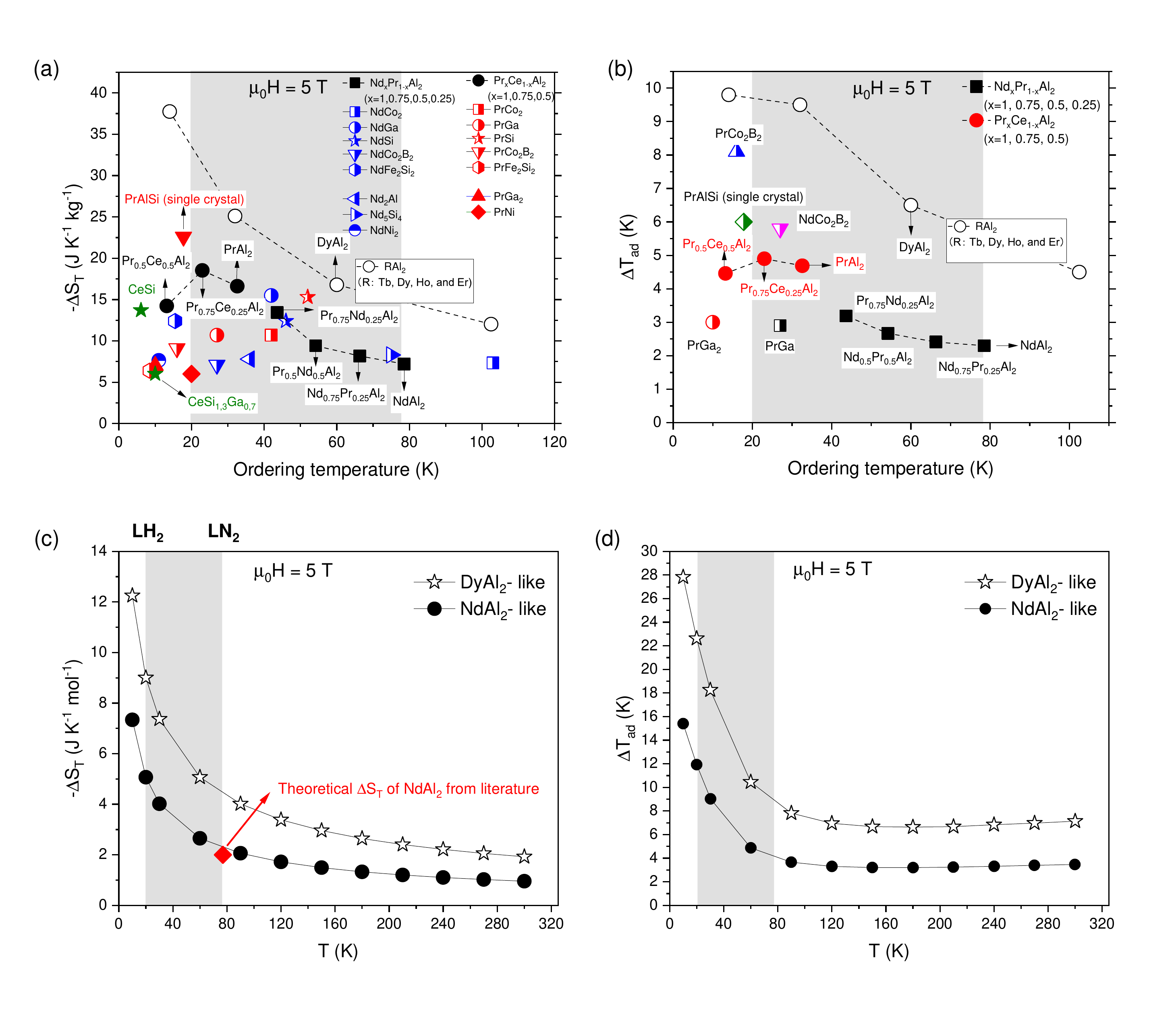}
    \caption{(a) (b) $\Delta S_T$ and $\Delta T_{ad}$ of light rare-earth-based intermetallic compounds. The heavy rare-earth-based \ce{RAl2} series is used for comparison. The light grey shadows highlight the temperature range of 77 $\sim$ 20 K. Data are taken from \cite{liu2023designing,zhang_magnetic_2015,dong_effect_2019,wang_magnetic_2014,zhang_magnetic_2015-1,zheng_nearly_2014,zheng_large_2018,Wang.2014,Paixao.2020,Plaza.2009,Li.2011c,Li.2009,Ma.2019b,Pecharsky.2003,Lyu.2020, Liu.2022,murtaza_magnetocaloric_2020,synoradzki_magnetocaloric_2022,dos_reis_study_2010,kumar_magnetism_2008} (seen in supplementary). (c) (d) $\Delta S_T$ and $\Delta T_{ad}$ of \ce{DyAl2}-like ($g_J = 4/3$, $J = 7.5$, and $T_D =384 K$) and \ce{NdAl2}-like ($g_J = 8/11$, $J = 4.5$ and $T_D =384 K$) series from mean-field approach.}
    \label{Fig8}
\end{figure*}

\textbf{\Cref{Fig8}} (a) plots the $\Delta S_T$ of the light rare-earth-based intermetallic compounds. The heavy rare-earth-based Laves phases \ce{RAl2} (R = Tb, Dy, Ho, and Er) serve as references for comparing the MCEs of the light and heavy rare-earth-based MC materials, as they are regarded as good MC materials for hydrogen liquefaction due to their excellent $\Delta S_T$ and $\Delta T_{ad}$ in the temperature range of 77 $\sim$ 20 (K) \cite{Liu.2022,Khan.2010}. Consistent with the conclusion made in \textbf{\Cref{increasingtrend}}, an increasing trend of $\Delta S_T$ with the decreasing $T_C$ is observed for light rare-earth-based MC materials. In addition, light rare-earth-based intermetallic compounds can also achieve excellent values of $\Delta S_T$ near the hydrogen condensation point (20 K). In magnetic fields of 5 T, a single crystal \ce{PrAlSi} exhibits a $\Delta S_T$ of \qty{22.6}{\joule\per\kelvin\per\kilogram} \cite{Lyu.2020}, being very close to the $\Delta S_T$ of \ce{HoAl2} which is regarded as competitive candidate for the final stage of hydrogen liquefaction \cite{Khan.2010}. The excellent values of $\Delta S_T$ are also achieved in \ce{(Pr,Ce)Al2} near \qty{20}{\kelvin} under a magnetic field change of \qty{5}{\tesla} \cite{liu2023designing}. All the \ce{(Pr,Ce)Al2} samples show a $\Delta S_T$ that is larger than or comparable to \ce{DyAl2} with a competitive $\Delta S_T$ for the initial stage of hydrogen liquefaction \cite{Liu.2022}. At \qty{52}{K}, which is a temperature in the middle stage of hydrogen liquefaction, large $\Delta S_T$  can still be observed in \ce{PrSi} that is comparable to \ce{DyAl2} \cite{wang_magnetic_2014,Liu.2022}.  \par

\textbf{\Cref{Fig8}} (b) plots the $\Delta T_{ad}$ of the light rare-earth-based intermetallic compounds with the heavy rare-earth series \ce{RAl2} (R = Tb, Dy, Ho, and Er) serving as a reference for comparison \cite{Liu.2022}. Although the data of $\Delta T_{ad}$ are not as rich as that of $\Delta S_T$, an increasing trend of $\Delta T_{ad}$ with the decreasing $T_C$ is still observable, which agrees with the conclusion made in \textbf{\Cref{increasingtrend}}. The \ce{PrCo2B2} exhibits the largest $\Delta T_{ad}$ of about 8 K at 16 K, only \qty{2}{K} smaller than \ce{HoAl2} \cite{Li.2011c,Liu.2022}. The light rare-earth-based materials can also show excellent MCEs in the vicinity of the hydrogen condensation point, making them appealing for the final stage of MC hydrogen liquefaction. \par
 
 Furthermore, two performance gaps can be summarized from \textbf{\Cref{Fig8}} (a) and (b): (1)  both $\Delta S_T$ and $\Delta T_{ad}$ of the light rare-earth-based intermetallic compounds show a smaller value than that of their heavy rare-earth-based counterparts with a similar $T_C$; (2) both $\Delta S_T$ and $\Delta T_{ad}$ of the rare-earth-based intermetallic compounds generally show a smaller value near the nitrogen condensation point than near the hydrogen condensation point. As shown in \textbf{\Cref{Fig8}} (a) and (b), all values of $\Delta S_T$ and $\Delta T_{ad}$ of light rare-earth-based intermetallic compounds are below the curves of the heavy rare-earth-based Laves phase series \ce{RAl2}.  Moreover, compounds such as \ce{DyAl2} with a $T_C$ closer to 77 K show smaller $\Delta S_T$ and $\Delta T_{ad}$ than compounds such as \ce{HoAl2} with a $T_C$ closer to 20 K. It is the same case when comparing \ce{PrAl2} with a $T_C$ closer to 20 K and \ce{NdAl2} with a $T_C$ closer to 77 K.   \par

These two performance gaps can be well explained by mean-field theory. \textbf{\Cref{Fig8}} (c) and (d) compare the $\Delta S_T$ and $\Delta T_{ad}$ of a \ce{NdAl2}-like series with a \ce{DyAl2}-like series. Since crystalline electrical field can also influence $\Delta S_T$ \cite{de_oliveira_theoretical_2010}, in \textbf{\Cref{Fig8}} (c) the theoretical $\Delta S_T$ from reference \cite{carvalho_experimental_2005} which takes the crystalline electrical field into consideration is also included. This theoretical value is near the profile of the \ce{NdAl2}-like series. Agreeing with the observations in \textbf{\Cref{Fig8}} (a) and (b), both $\Delta S_T$ and $\Delta T_{ad}$ of the \ce{NdAl2}-like series calculated from mean-field approach are below the curves of the \ce{DyAl2}-like series. In addition, both $\Delta S_T$ and $\Delta T_{ad}$ show significantly larger values near the hydrogen condensation point than near the nitrogen condensation point. \par

Despite the relatively low resource criticality of light rare-earth elements, which makes them attractive for large-scale applications, the less strong MCE near 77 K questions the feasibility of using the light rare-earth-based second-order MC materials for cooling hydrogen gas (precooled by liquid nitrogen) from 77 K to lower temperature. \par

\subsubsection{Substituting highly critical rare earths}

\begin{figure}[ht]
	\centering
    \includegraphics[width = \linewidth]{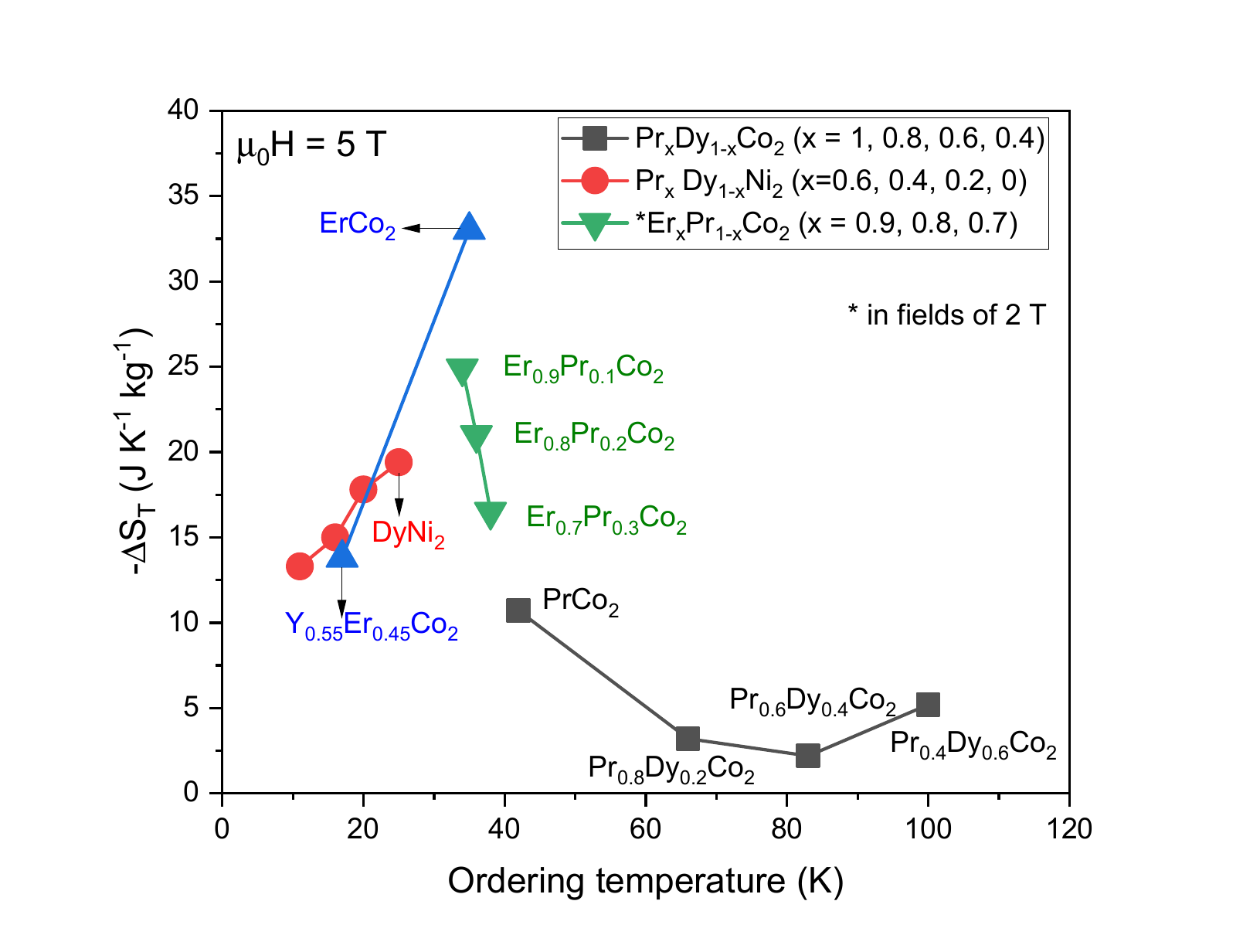}
    \caption{$\Delta S_T$ of MC intermetallic compounds of \ce{(R1, R2)X2} where \ce{X} is a non-rare-earth element and \ce{R1} and \ce{R2} are the heavy rare-earth elements and the less critical rare-earth elements. Data are taken from \cite{dong_effect_2019,Chen.2022,Baranov.2009,ParraBorderias.2009} (seen in supplementary).}
    \label{Fig9}
\end{figure}

A further compromise to the dilemma of the performance and criticality of rare-earth-based MC intermetallic compounds is partially substituting the more critical heavy rare-earth elements by the less critical rare-earth elements at the rare-earth sites. Due to their lower resource criticality, the Yttrium and light rare-earth elements La, Ce, Nd, and Pr are often chosen to replace the highly critical heavy rare-earth elements. \textbf{\Cref{Fig9}} presents the $\Delta S_T$ of the rare-earth-based MC intermetallic compounds by mixing highly critical heavy rare-earth elements and less critical rare-earth elements. In the cases of \ce{(Y,Er)Co2}, \ce{(Er,Pr)Co2},and \ce{(Pr,Dy)Co2}, the higher content of the heavy rare-earth elements, the greater the  $\Delta S_T$. This can be explained by the fact that the heavy rare-earth ions possess larger magnetic moments. However, in the case of  \ce{(Pr,Dy)Co2}, opposite trend is observed. \par

The coupling effect between light and heavy rare earths in the rare-earth-based intermetallic compounds has been investigated dating back to the 60s by Swift and Wallace \cite{swift_magnetic_1968}. For the \ce{(R,R{'})Al2} (R: light rare earths, R’: heavy rare earths), the spins are observed to couple ferromagnetically in most cases. An antiferromagnetic coupling was observed in the \ce{(Ce, Eu)Al2} series. It should be emphasized that the total angular momentums of light and heavy rare earths are all ferrimagnetically alignment for the \ce{(R,R{'})Al2} investigated in Swift and Wallace's article. It is worth mentioning that Swift and Wallace discovered that the saturation moments in many cases are less than expected from $g_JJ$, which suggests a high crystal quenching effects. In 2014, Pathak \textit{et al}. demonstrated that Pr and Er couple antiferromagnetically, resulting in a coexistence of ferromagnetism, ferrimagnetism, and meta-magnetism \cite{pathak_unexpected_2014}. More recently, Del Rose \textit{et al}. also observed the antiferromagnetic coupling in \ce{(Pr,Gd)ScGe} by XMCD \cite{del_rose_origins_2022}.

Generally speaking, mixing light- and heavy rare-earth elements is not intensively studied. Further investigations need to be done to understand the coupling between light rare-earth and heavy rare-earth ions. Nevertheless, from the above analysis, it can be concluded that mixing rare-earth elements with different resource criticality is an effective way to tune the ordering temperature and can reduce the rare-earth criticality to a certain extent. The challenge is to find an optimal balance of performance and criticality. \par

\subsection{High performance as well as low criticality}
The aforementioned difficulty in achieving high MC performance of second-order light rare-earth-based MC materials near the condensation point of nitrogen (77 K) can be overcome by developing light rare-earth-based MC materials with a first-order phase transition. This subsection focuses on light rare-earth-based first-order MC intermetallic compounds. In addition, the Gd-based MC intermetallic compounds are discussed, as Gd is the least critical heavy rare-earth elements in resources.

\subsubsection{Light rare-earth-based first-order MC materials}

\begin{figure*}[ht]
	\centering
    \includegraphics[width = \linewidth]{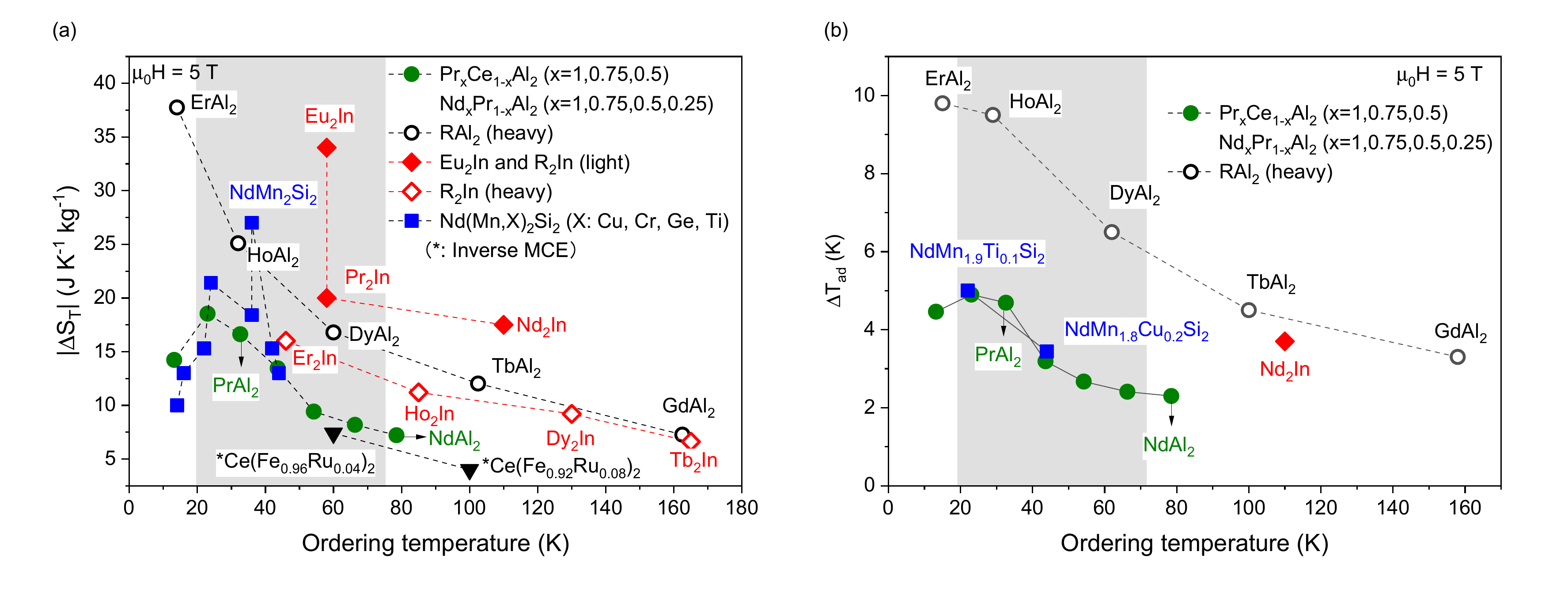}
    \caption{(a) (b) $\Delta S_T$ and $\Delta T_{ad}$ of light rare-earth based first-order MC materials. The light- and heavy rare-earth-based Laves phase series and \ce{Eu2In} are used for comparisons. Data are taken from \cite{liu2023designing,Liu.2022,Zhang.2011,Guillou.2018,Zhang.2009,Zhang.2009b,Zhang.2009c,md_din_magnetic_2013,md_din_magnetic_2014-1,md_din_magnetic_2014,din_magnetic_2014,chattopadhyay_magnetocaloric_2006,Biswas.2022,biswas_unusual_2022} (seen in supplementary). }
    \label{Fig10}
\end{figure*}

The discovery of excellent $\Delta S_T$ in \ce{Pr2In} and \ce{Nd2In} breaks the performance gaps between light- and heavy rare-earth-based intermetallic compounds and between MCEs near 20 K (hydrogen condensation point) and near 77 K (nitrogen condensation point) \cite{liu_large_2021,biswas_first-order_2020,biswas_unusual_2022,Biswas.2022}. \textbf{\Cref{Fig10}} (a) plots  $\Delta S_T$ of the selected light rare-earth-based first-order MC materials. In magnetic fields of \qty{5}{\tesla} and at \qty{57}{\kelvin}, \ce{Pr2In} exhibits a $\Delta S_T$ of about \qty{20}{\joule\per\kelvin\per\kilogram}, surpassing \ce{Er2In} known for showing the largest $\Delta S_T$ of about \qty{15.5}{\joule\per\kelvin\per\kilogram} among the heavy rare-earth-based \ce{R2In} (R= Gd, Tb, Dy, Ho, and Er) series \cite{Zhang.2011}.  \par

The impressive $\Delta S_T$ in \ce{Pr2In} is ascribed to the first-order magnetic phase transition at about 57 K \cite{biswas_first-order_2020,Biswas.2022}. Similar first-order magnetic phase transitions were also reported in \ce{Eu2In} \cite{Guillou.2018} and \ce{Nd2In} \cite{liu_large_2021,biswas_unusual_2022}. All of them show an outstanding $\Delta S_T$ that is not only larger than their heavy rare-earth counterparts (\ce{R2In}), but also the heavy rare-earth-based \ce{DyAl2}. \par

Another eye-catching feature of \ce{Pr2In}, \ce{Nd2In}, and \ce{Eu2In} is that unlike most first-order MC materials that exhibit significant thermal hysteresis, these three first-order MC materials demonstrate negligible thermal hysteresis, as reported by Guillou \textit{et al.} for \ce{Eu2In} \cite{Guillou.2018}, Biswas \textit{et al.} for \ce{Pr2In} \cite{biswas_first-order_2020}, and Liu \textit{et al.} for \ce{Nd2In} \cite{liu_large_2021}. The negligible thermal hysteresis might be related to their small volume changes during the first-order phase transitions, as reported in reference \cite{liu_large_2021}. \par 

Although \ce{Pr2In}  demonstrates that light rare-earth-based MC materials can also achieve excellent $\Delta S_T$ near the nitrogen condensation point, more research on this compound and its relatives \ce{Eu2In} and \ce{Nd2In} are needed. Since their MCEs were just discovered within the last five years, their peculiar phase transitions are not well understood. Although demonstrated by Biswas \textit{ et al.} and Tapia-Mendive \textit{et al.} that the mechanism of the first-order phase transitions in \ce{Pr2In}, \ce{Nd2In}, and \ce{Eu2In} might be due to a fully electronic origin \cite{Biswas.2022,biswas_unusual_2022,MendiveTapia.2020}, second-order phase transitions in \ce{Pr2In} and \ce{Nd2In} were also reported \cite{cui_unconventional_2022}. The decomposition of the surfaces of the  \ce{Pr2In}, \ce{Nd2In}, and \ce{Eu2In} samples makes the investigations on their microstructures difficult \cite{biswas_first-order_2020,Guillou.2018,liu_large_2021}. However, microstructure is of great importance for exploring the nature of the phase transition, and these works need to be done in the near future. It should be emphasized that indium is also a highly critical element, and more work needs to be done to replace it. \par

Besides the light rare-earth based \ce{R2In} material system, the \ce{Nd(Mn,X)2Si2} is another first-order MC material system which is based on the rock-forming elements (Mn and Si) and shows excellent $\Delta S_T$ within the temperature range of 77 $\sim$ 20 K \cite{md_din_magnetic_2013,md_din_magnetic_2014-1,md_din_magnetic_2014,din_magnetic_2014}. In magnetic fields of 5 T, the $\Delta S_T$ of \ce{NdMn2Si2} reaches \qty{27}{\joule\per\kelvin\per\kilogram} at \qty{36}{\kelvin} \cite{md_din_magnetic_2013}, even surpassing one of the best heavy rare-earth-based MC intermetallic compound \ce{HoAl2} \cite{Liu.2022}. However, after doping the Mn site with other 3d elements to tailor the ordering temperature,  $\Delta S_T$ decreases significantly from \qtyrange[range-units = single]{27}{10}{\joule\per\kelvin\per\kilogram} \cite{md_din_magnetic_2013}. The $\Delta T_{ad}$ of \ce{NdMn_{1.9}Ti_{0.1}Si2} and \ce{NdMn_{1.8}Cu_{0.2}Si2} are comparative to the light rare-earth Laves series \cite{md_din_magnetic_2013,md_din_magnetic_2014-1}. \par

Ce-based first-order MC materials are not as intensively studied as the Nd- and Pr-based ones because of their relatively poor MCEs. However, considering that Ce is the most abundant rare-earth element, a Ce-based MC material with large $\Delta S_T$ and $\Delta T_{ad}$ would be of great importance.  The \ce{Ce(Fe_{0.96}Ru_{0.04})2} is a first-order MC materials with an inverse MCE, showing a value of $\Delta S_T$ of about \qty{7.4}{\joule\per\kelvin\per\kilogram} at about \qty{60}{\kelvin} \cite{chattopadhyay_magnetocaloric_2006}. Obviously, Ru is also a highly critical element. Nevertheless, \ce{Ce(Fe_{0.96}Ru_{0.04})2} sets an example of developing Ce-based first-order MC materials. \par

 \begin{figure}[ht]
	\centering
    \includegraphics[width = \linewidth]{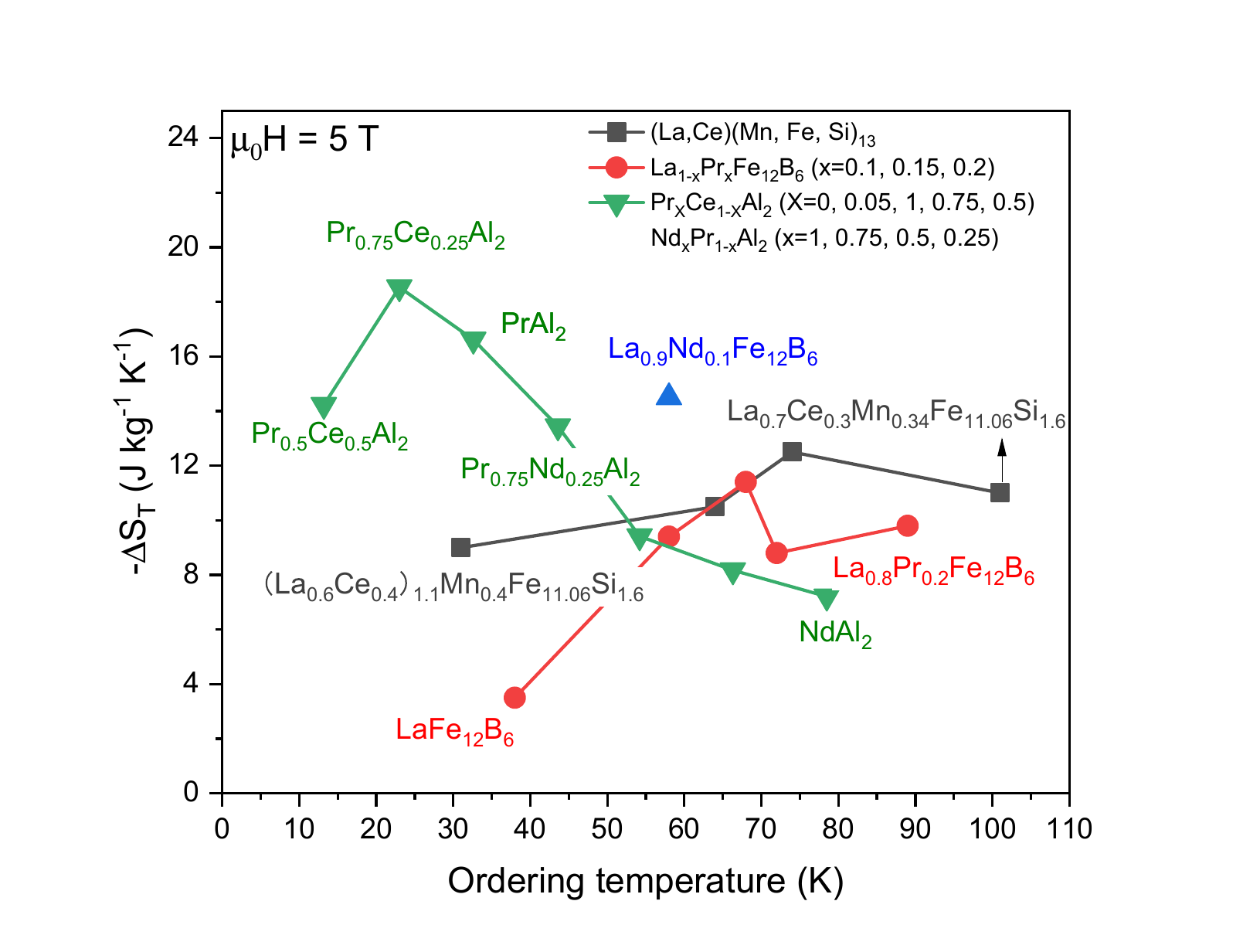}
    \caption{$\Delta S_T$ of La-based first-order MC intermetallic compounds. The light rare-earth Laves phases \ce{(R,R^{'})Al2} are used for making a comparison . Data are taken from \cite{Diop.2016,ma_achievement_2021,lai_reduction_2021} (seen in supplementary).}
    \label{Fig11}
\end{figure}

Another light rare-earth-based material family that exhibits first-order MCE is the La- based intermetallic compounds, namely \ce{La(Fe,Si)13} \cite{lai_reduction_2021} and \ce{LaFe12B6} \cite{Diop.2016,chen_enhancing_2021, ma_achievement_2021}. \textbf{\Cref{Fig11}} plots the $\Delta S_T$ of these two systems. It is worth mentioning that the nature of the phase transition order of \ce{La(Fe,Si)13} depends on the composition. In this part, we focus on the \ce{La(Fe,Si)13} compounds that exhibit first-order phase transitions. \ce{La(Fe,Si)13} is one of the most studied MC material systems due to its high potential to be used for near-room-temperature MC refrigeration \cite{Gutfleisch.2016,benke_magnetic_2020,moreno-ramirez_role_2018,moreno-ramirez_tunable_2019,shao_high-performance_2017,terwey_influence_2020,fujita2003itinerant,hu_direct_2003}. A recent study revealed that the ordering temperature of \ce{La(Fe,Si)13} can be further tuned to about 31 K by doping Ce and Mn, demonstrating that this material system has the potential to be used for MC hydrogen liquefaction \cite{lai_reduction_2021}. \par

In addition to \ce{La(Fe,Si)13}, the \ce{LaFe12B6} material system has caught the attention of researchers due to its first-order phase transition near 50 K. However, due to the fact that \ce{LaFe12B6} remains antiferromagnetic below 5 T, significant $\Delta S_T$ only appears in sufficiently high magnetic fields \cite{Diop.2016}. To make this material system useful for MC hydrogen liquefaction under a practical magnetic field change, its critical magnetic field needs to be reduced. Recent studies show that doping the La site with Pr or Nd transforms the antiferromagnetic to ferromagnetic configuration \cite{chen_enhancing_2021,ma_achievement_2021}. The MCE is greatly enhanced, from about \qty{5}{\joule\per\kelvin\per\kilogram} for \ce{LaFe12B6} in magnetic fields of \qty{5.5}{\tesla} up to \qty{11.4}{\joule\per\kelvin\per\kilogram} for \ce{La_{0.9}Pr_{0.1}Fe12B6} \cite{ma_achievement_2021} and \qty{14.5}{\joule\per\kelvin\per\kilogram} for \ce{La_{0.9}Nd_{0.1}Fe12B6} \cite{chen_enhancing_2021} in magnetic fields of 5 T. \par

Although the La-based first-order MC materials have great potential, their performances near 20 K need to be improved. Compared to the light rare-earth-based Laves phase series, the La- based first-order MC materials exhibit a significant larger $\Delta S_T$ in the vicinity of 77 K, but a much smaller $\Delta S_T$ near 20 K. In addition, their reversibility needs further investigation as they show significant thermal hysteresis \cite{lai_reduction_2021,Diop.2016,chen_enhancing_2021,ma_achievement_2021}. In particular, the thermal hysteresis of the \ce{LaFe12B6} material system is about 20 K \cite{Diop.2016}. \par

From the analysis above, it can be concluded that light rare-earth-based first-order MC materials can overcome the disadvantage of the light rare-earth-based second-order MC materials by exhibiting large $\Delta S_T$ near 77 K. First-order light rare-earth-based intermetallic compounds \ce{Nd2Fe2Si2} and \ce{Pr2In} even present a $\Delta S_T$ that surpasses the heavy rare-earth \ce{RAl2} materials with similar ordering temperatures. However, the reversibility of light rare-earth-based MC materials needs further investigation.

\subsubsection{Gd-based MC materials}

Although light rare-earth elements have the advantage of being less critical in resources, the heavy rare-earth elements still hold one overwhelming advantage of exhibiting much larger magnetic moments.  Gd shows an effective magnetic moment of about \qty{7.94}{\mu_B}, more than two times larger than Pr and Nd with an effective magnetic moment of 3.58 and \qty{3.52}{\mu_B} respectively \cite{coey2010magnetism}. \par

Gd is the most abundant heavy rare-earth element \cite{liu2023designing}. In fact, its oxides are also as cheap as the oxides of Nd and Pr \cite{liu2023designing}. Although Nd and Pr are much more abundant than Gd, the rare-earth permanent magnet industry consumes a large amount of Nd and Pr metals, while Gd is less widely used in industry. The large magnetic moment of Gd and the relatively cheap price of its oxides make its alloys a competitive candidate with high cost-effectiveness for MC hydrogen liquefaction. \par

\begin{figure}[ht]
	\centering
    \includegraphics[width = \linewidth]{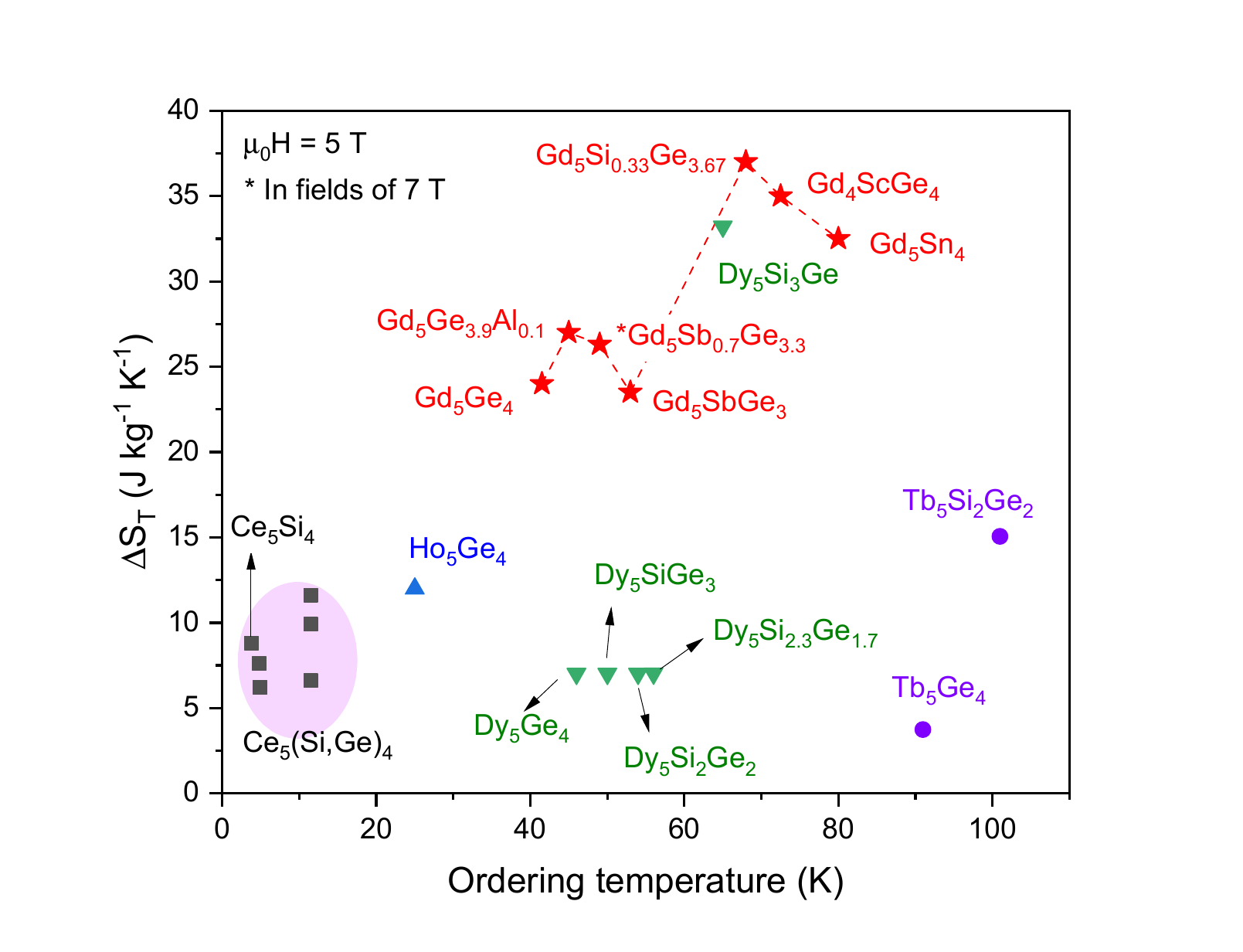}
    \caption{$\Delta S_T$ of rare-earth-based \ce{R5T4} (T: non-rare-earth elements) compounds. Data are taken from \cite{gschneidner_nonpareil_2000,arnold_pressure_2009,zhang_phase_2010,chernyshov_structural_2006,mudryk_enhancing_2017,morellon_magnetocaloric_2001,huang_preparation_2002,balachandran_magnetothermal_2000,nirmala_metamagnetism-enhanced_2016,Ryan.2003,zou_ferromagnetic_2013} (seen in supplementary).}
    \label{Fig12}
\end{figure}

One of the milestones of magnetic cooling is the discovery of the giant room-temperature MCE in \ce{Gd5Si2Ge2} by Pecharsky \textit{et al.} in 1997 \cite{Pecharsky.1997b}. In the same year, \ce{Gd5(Si,Ge)4} was reported to also show giant MCE at cryogenic temperatures \cite{pecharsky_tunable_1997}. Since then, many rare-earth-based \ce{R5T4} (T: non-rare-earth element) have been discovered to show a significant MCE. \textbf{\Cref{Fig12}} shows the $\Delta S_T$ of the rare-earth-based \ce{R5T4} compounds with an ordering temperature in the cryogenic temperature range. Comparing to the other \ce{R5T4} compounds, the \ce{Gd5(Si, Ge)4} \cite{gschneidner_nonpareil_2000,arnold_pressure_2009}, \ce{Gd5Ge_{3.9}Al_{0.1}} \cite{zou_ferromagnetic_2013}, \ce{Gd5SbGe3} \cite{chernyshov_structural_2006}, \ce{Gd4ScGe4} \cite{mudryk_enhancing_2017}, and \ce{Gd5Sn4} \cite{Ryan.2003} stand out for showing the largest values of $\Delta S_T$ in the vicinity of their ordering temperatures. One of the disadvantages of using \ce{Gd5(Si,Ge)4} for MC hydrogen liquefaction is that Ge also belongs to high critical elements \cite{Gauss2021rare}. It would be of great importance to reduce or replace Ge with other less critical elements. It is worth mentioning that the magnetic properties of \ce{Gd5(Sb,Ge)4} \cite{chernyshov_magnetostructural_2009,chernyshov_structural_2006} and \ce{Gd5Ge_{4-x}P_{x}} (x=0.25–0.63) \cite{cheung_structural_2013} have been studied, which have shown or indicated that they could be good MC materials in terms of performance at cryogenic temperatures. 

 \begin{figure}[ht]
	\centering
    \includegraphics[width = \linewidth]{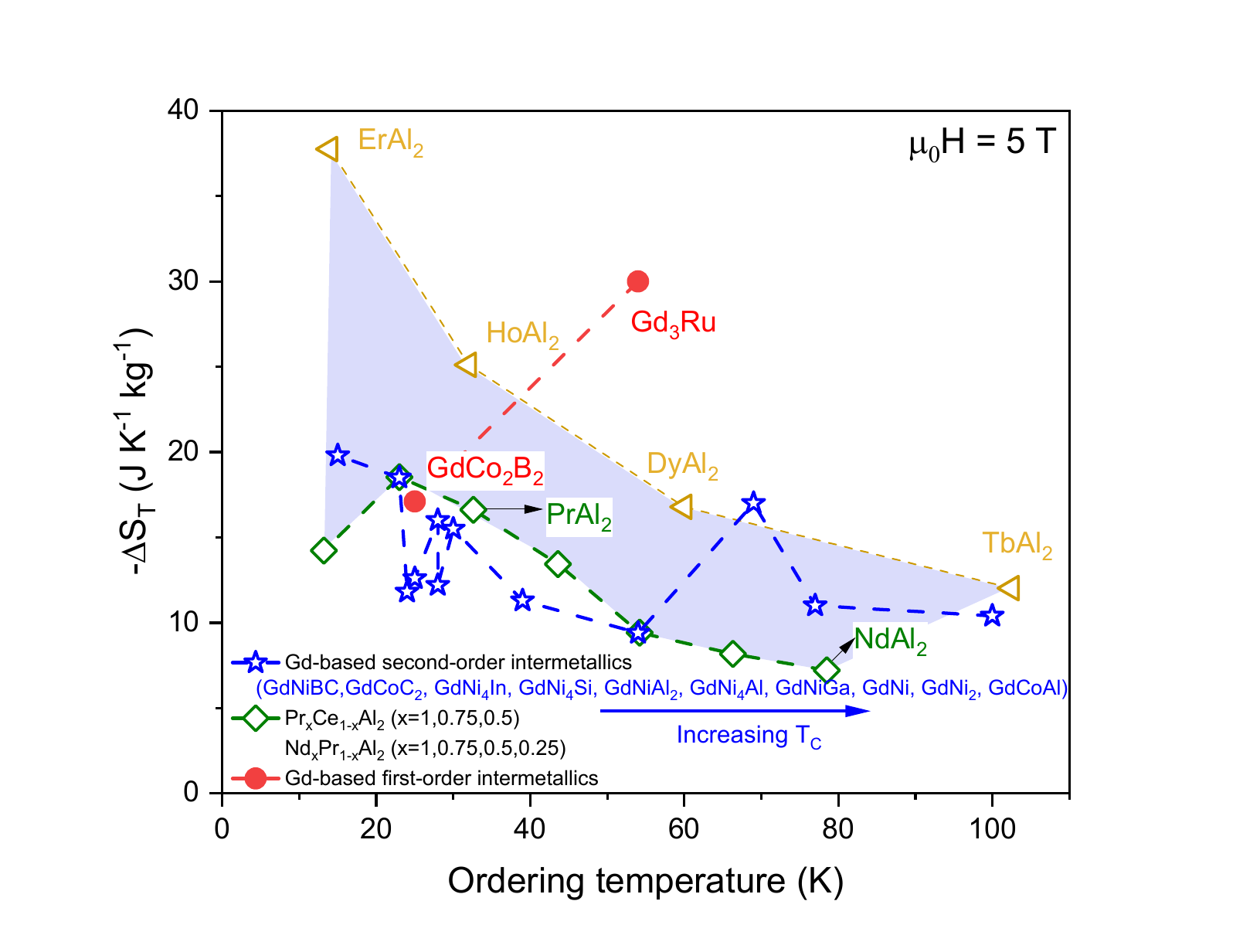}
    \caption{$\Delta S_T$ of Gd-based MC intermetallic compounds. The light and heavy \ce{RAl2} Laves phases serving as a reference. The light blue shadow marks the area between light and heavy \ce{RAl2} Laves phases. Data are taken from \cite{Rajivgandhi.2017,Taskaev.2020,meng_magnetic_2016,dembele_large_2015,tolinski_magnetocaloric_2012,peng_magnetic_2022,tolinski_magnetic_2002,zhang_magnetic_2001,monteiro_physical_2015,li_giant_2009} (seen in supplementary).}
    \label{Fig13}
\end{figure}

\textbf{\Cref{Fig13}} displays the $\Delta S_T$ of the Gd-based MC materials (excluding the\ce{Gd5T4} compounds), with the light- and heavy rare-earth series \ce{RAl2} (R: Ce, Pr, Nd, Tb, Dy, Ho, and Er) plotted for comparison. Gd-based MC materials show both excellent $\Delta S_T$ near 20 K and near 77 K. Though belonging to second-order MC materials, \ce{GdCo2C2} has a $\Delta S_T$ of \qty{27.4}{\joule\per\kelvin\per\kilogram} at 15 K \cite{meng_magnetic_2016}, and \ce{GdNi} shows a $\Delta S_T$ of \qty{17.3}{\joule\per\kelvin\per\kilogram} at \qty{69}{\kelvin} \cite{Rajivgandhi.2017}. It should be emphasized that Ru belong to high critical elements as well, but \ce{Gd3Ru} sets an example of Gd-based first-order MC materials showing excellent $\Delta S_T$ near the condensation point of nitrogen. \par

It is worth mentioning that if Gd is used in large amounts, the Gd resource would become more critical. It has been revealed that Gd cannot sustain a mass-market MC refrigeration and air conditioning market \cite{Gauss2021rare}. The size of the market that Gd can support for MC hydrogen liquefaction needs further investigation. 

\section{Conclusions}

This review focuses on the rare-earth-based MC intermetallic compounds for hydrogen liquefaction. The feature that second-order MC materials can show the same excellent $\Delta S_T$ and $\Delta T_{ad}$ at low temperatures as the typical giant first-order MC materials is revealed. The increasing trends of the maximum $\Delta S_T$ and $\Delta T_{ad}$ with the decreasing $T_C$ are summarized by reviewing the heavy rare-earth-based magnetocaloric intermetallic compounds.  \par

Using a mean-field approach, the increasing trends and the "giant" values of $\Delta S_T$ and $\Delta T_{ad}$ at low temperature are theoretically demonstrated. The increasing trend of $\Delta S_T$ as $T_C$ decreases and the "giant" $\Delta S_T$ at low temperature can be well understood by the power law of $\Delta S_T(T_C) \propto T_C^{-2/3}$. The influence of cryogenic temperature on the MC performance is ascribed to the weaker thermal motion of magnetic moments and the strongly decreasing heat capacity in the cryogenic temperature range. \par

Since MCEs of rare-earth-based intermetallic compounds are highly correlated to their $T_C$, methods to tailor the MCEs of rare-earth-based intermetallic compounds by tuning the $T_C$ to cover the temperature range of 77 $\sim$ 20 K required by MC hydrogen liquefaction are reviewed. Three methods are summarized: (1) mixing the rare-earth elements with different de Gennes factors at the rare-earth sublattice sites; (2) doping the non-rare-earth sublattice sites; (3) hydrogenation.

The last part of this article focuses on resolving the dilemma of performance and criticality. Reviewing also the light rare-earth-based second-order MC intermetallic compounds, it is demonstrated that they can achieve excellent MCEs in the vicinity of the hydrogen condensation point (20 K). However, a bottleneck is revealed that near the nitrogen condensation point (77 K), light rare-earth-based second-order MC materials show less strong MCEs. A method to decrease the rare-earth criticality of heavy-rare-earth-based MC intermetallic compounds is discussed: substituting the heavy rare-earth elements with less critical light rare-earth elements and yttrium. By partially sacrificing the MC performances, the criticality can be reduced.\par

By also assessing the light rare-earth-based first-order MC intermetallic compounds, we show that both a high MC performance and a low rare-earth criticality can be obtained. \ce{NdMn2Si2} and \ce{Pr2In} exhibit a comparable $\Delta S_T$ to the materials with a similar ordering temperature in the heavy rare-earth-based \ce{RAl2} series. The La- based first-order MC materials, namely \ce{La(Fe,Si)13} and \ce{LaFe12B6}, show high potential to be used for MC hydrogen liquefaction. However, their reversibility needs more investigation. Since Gd is the least critical heavy rare-earth metal in resources, its alloys for MC hydrogen liquefaction are particularly reviewed. Both of the second-order and first-order Gd-based MC materials show a competitive MCE within temperature 77 $\sim$ 20 K, overcoming the disadvantage of the small $\Delta S_T$ of the second-order light rare-earth-based MC materials near 77 K. \par

In summary, this review reveals that operating at cryogenic temperature has a positive effect on MC performance, while the high criticality of heavy rare-earth elements has a negative effect on the economic viability of using rare-earth-based intermetallic compounds for large-scale applications of MC hydrogen liquefaction. In addition, ways to mitigate the criticality of rare-earth elements while optimizing the MCE have been explored. \par

\section*{Data Availability}

The data that support the findings of this study are available from the corresponding author upon reasonable request.

\section*{Acknowledgement}
This article is dedicated to the memory of Prof. Dr. Vitalij Pecharsky, who was a distinguished scientist known for his pioneering work in materials science and magnetic cooling. His legacy in research and education will continue to inspire future generations of scientists, in particular those from the MCE community.

This work is supported by the CRC/TRR 270 (Project-ID 405553726 and 456263705), the ERC under the European Union's Horizon 2020 research and innovation program (Grant No. 743116, Cool Innov), the Clean Hydrogen Partnership and its members within the framework of the project HyLICAL (Grant No. 101101461), the Helmholtz Association via the Helmholtz-RSF Joint Research Group (Project No. HRSF-0045), and the HLD at HZDR (member of the European Magnetic Field Laboratory (EMFL)). 

In addition, we thank Xing Tang from National Institute of Materials Science (NIMS), Japan, for his valuable discussions and advice.



\printbibliography

\end{document}